\documentclass[%
preprintnumbers,
 amsmath,amssymb,
 aps,
]{revtex4-2}

\usepackage{graphicx}% Include figure files
\usepackage{dcolumn}% Align table columns on decimal point
\usepackage{bm}% bold math
\usepackage{color,graphicx}% color
\usepackage[dvipsnames]{xcolor}
\usepackage{ulem}

\usepackage{epstopdf, epsfig}
\usepackage{subcaption}
\usepackage{dashrule}
\usepackage{multirow}
\usepackage{natbib,url,hyperref}
\usepackage{amsmath}

\usepackage{hyperref}% add hypertext capabilities

\definecolor{grey}{rgb}{0.7,0.7,0.7}

\definecolor{cNeutralGray}{RGB}{99,101,105}
\definecolor{cGreenCurrent}{rgb}{0.4660,0.6740,0.1880}
\definecolor{cEverGreen}{rgb}{0,0.5,0}
\definecolor{cCyanCurrent}{rgb}{0.3010, 0.7450, 0.9330}
\definecolor{cBlueSecondary}{RGB}{0, 75, 135}
\definecolor{cViolet}{RGB}{204, 0, 204}
\definecolor{cYellow}{RGB}{255, 255, 51}
\definecolor{cOrangeUtility}{RGB}{242, 160, 0}
\definecolor{cRedUtility}{RGB}{183, 49, 44}

\definecolor{cBlueBrand}{RGB}{47, 126, 178}
\definecolor{cVioletCurrent}{rgb}{0.4940, 0.1840, 0.5560}
\definecolor{cBlack}{rgb}{0, 0, 0}
\definecolor{cYellowCurrent}{rgb}{0.9290, 0.6940, 0.1250}

\begin{document}

% \preprint{APS/123-QED}

% \title{High-resolution LES of a simplified urban environment}% Force line breaks with \\
%\thanks{A footnote to the article title}%f
\title{High-resolution large-eddy simulations of simplified urban flows}

\author{Marco Atzori}
 \email{marco.atzori@jku.at}
 \affiliation{%
 Department of Particulate Flow Modelling, Johannes Kepler University, Linz 4040, Austria
}%

\author{Pablo Torres}%
\affiliation{%
 Instituto Universitario de Matem\'atica Pura y Aplicada, Universitat Polit\`ecnica de Val\`encia, Valencia 46022, Spain
}%

\author{Alvaro Vidal}%
\affiliation{%
 Parallel Works, Chicago IL 60654, USA
}%

\author{Soledad Le Clainche}%
\affiliation{%
 School of Aerospace Engineering, Universidad Polit\`ecnica de Madrid, Madrid 28040, Spain
}%

\author{Sergio Hoyas}%
\affiliation{%
 Instituto Universitario de Matem\'atica Pura y Aplicada, Universitat Polit\`ecnica de Val\`encia, Valencia 46022, Spain {\color{white} a}
}%

\author{Ricardo Vinuesa}%
\affiliation{%
 FLOW, Engineering Mechanics, KTH Royal Institute of Technology, Stockholm 10044, Sweden
}%

\date{\today}% It is always \today, today,

\begin{abstract}
High-fidelity large-eddy simulations of the flow around two rectangular obstacles are carried out at a Reynolds number of $10,000$ based on the free-stream velocity and the obstacle height. The incoming flow is a developed turbulent boundary layer. Mean-velocity components, turbulence fluctuations, and the terms of the turbulent-kinetic-energy budget are analyzed for three flow regimes: skimming flow, wake interference, and isolated roughness. Three regions are identified where the flow undergoes the most significant changes: the first obstacle's wake, the region in front of the second obstacle, and that around the second obstacle. In the skimming-flow case, turbulence activity in the cavity between the obstacles is limited and mainly occurs in a small region in front of the second obstacle. In the wake-interference case, there is a strong interaction between the free-stream flow that penetrates the cavity and the wake of the first obstacle. This interaction results in more intense turbulent fluctuations between the obstacles. In the isolated-roughness case, the wake of the first obstacle is in good agreement with that of an isolated obstacle. Separation bubbles with strong turbulent fluctuations appear around the second obstacle.
\end{abstract}


% TEXT FOR TEASER (A TEASER WILL BE NEEDED, and a figure)
%
% FIGURE FOR TEASER IS wing2bis.png
\maketitle

%\tableofcontents

\section{Introduction}\label{intro}

Urban areas are a vital element of our society: currently, about 75\% of the population lives in cities in the European Union (EU), and it is estimated that by 2050, seven out of every ten people in the world will become urban residents \citep{UNO20}. Approximately 90\% of the urban population in the EU was exposed to air pollution levels that exceeded the levels recommended by the World Health Organization (WHO). Pollution leads to around 800,000 premature deaths in Europe every year~\citep{EEA19,lelieveld_et_al}. Moreover, sustainable cities are the eleventh Sustainable Development Goal of the United Nations. Therefore, there is an urgent need to improve forecasting and assessment methods to meet these challenges and achieve urban sustainability soon.

Due to the reasons above, the flow around building-like obstacles has been extensively studied \citep{isy78,bri79,zaj11} to improve pollutant dispersion, heat propagation, or energetic efficiency. For a complete review of these methods, we refer to Ref. \cite{tor21}. These studies are mainly based on empirical observations, meteorologic models, or experimental results. However, turbulence is present in a wide variety of physical phenomena, and urban environments are certainly no exception~\citep{oke88}. In this work, we present a numerical study to analyze the interaction of a developed turbulent boundary layer with two buildings in three different configurations.

Overall, the studies dealing with urban flows can be gathered around three main lines of investigation: experimental, numerical, and data-driven. Experimental studies tend to combine empirical descriptions of the flow with specific physical quantities measurements that are relevant to analyzing the flow dynamics. These kinds of works are usually divided by their scope: on the one hand, we find studies that characterize the overall dynamics of the flow in urban environments. Within this group, Oke \cite{oke88} observed that three zones of disturbance could characterize the flow in the envelope of a squared cross-section obstacle: ahead of the obstacle, a bolster eddy vortex; behind, a lee eddy that is drawn into a cavity of low pressure. Finally, a wake region appears downstream, characterized by increased turbulent intensity but lower horizontal speeds. In this way, fully understanding urban flows inevitably leads to the study of turbulence. We can also mention the work of Britter \& Hanna \cite{bri03}, describing the urban environment in terms of the length scale. These authors divided the urban environment into a wide range of scales bounded by the regional (from 10 to 200 km) and neighborhood (from 100 m to 2 km) scales. The present work deals only with the latter.

On the other hand, we find experimental studies that focus on particular flow applications. For instance, Di Sabatino \textit{et al.}~\cite{sab09} carried out the Phoenix Urban Heat Island experiment, in which an extensive database of temperature measurements in various areas of Central Phoenix, Arizona, was gathered. The authors used this database to study the urban heat island (UHI) in Central Phoenix and validate UHI models. Similarly, Weerasuriya \textit{et al.}~\cite{wee18} assessed the effect of twisted winds on pedestrian comfort. A scaled model of Tsuen Wan street in Hong Kong was tested in a wind tunnel to obtain the mean flow, turbulent intensities, and yaw angles. Pedestrian comfort was also studied experimentally by Corke \textit{et al.}~\cite{corke_et_al}, and pollutant dispersion was assessed in a number of urban environments by Nagib \& Corke \cite{nagib_corke} and Monnier \textit{et al.}~\cite{monnier_et_al}. More complicated geometries were also analyzed experimentally by Monnier \textit{et al.}~\cite{monnier_et_al2}.

Works based on experimental methodologies have proven helpful in studying local and applied phenomena. However, when dealing with a general scope, \textit{i.e.} focusing on the physics defining the flow behavior, experimental studies happen to fall short. That is why many studies use numerical simulations to characterize the overall dynamics of urban flows. There is a wide range of numerical methods available with different levels of accuracy. Since urban flows are highly complex, only high-fidelity methods can represent the flow fields with the highest level of detail. Reynolds-averaged Navier--Stokes (RANS) models in which all turbulent scales are modeled are not able to fully characterize the physical processes that take place in urban environments \citep{fer10,vit20}. Hence, the complexity of urban flows requires direct numerical simulations (DNS) or well-resolved large-eddy simulations (LES) to represent the flow fields faithfully. The studies using LES include Belcher \cite{bel05}, and Branford \textit{et al.}~\cite{bra11}, which incorporated passive scalars to simulate the dispersion of pollutants in urban environments. Also, Nakayama \textit{et al.}~\cite{nak11,nak12} ran an LES using the Smagorinsky model to simulate the flow due to solid wings in Tokyo. Other studies chose to use DNS to solve the flow fields. For example, Coceal \textit{et al.}~\cite{coc07} performed a DNS to obtain the turbulent statistics of the flow over an idealized urban environment. Vinuesa \textit{et al.}~\cite{vin15} ran a DNS of a wall-mounted square cylinder under laminar and turbulent inflows to characterize the flow in a single-obstacle environment.
More recently, Zhao \textit{et al.}~\cite{zhao_et_al_pof} studied the flow between two building-like obstacles for a laminar flow and several configurations. One idea developed by Torres \textit{et al.}~\cite{tor21} is to study the coherent structure existing in the flow in order to be able to formulate effective strategies to mitigate contamination. Here, {\it coherent structures} are defined as a three-dimensional flow region that satisfies a specific attribute—for example,  rotation or wave level. These structures are responsible for the transfer of momentum within the city and, therefore, the dispersion of pollutants.

Note that even if there are several studies on simulations of the flow in urban environments, the vast majority of these studies tend to focus on analyzing the mean velocities of the flow, thus leaving out important information on the nature of the flow. For example, Refs. \cite{ger91,coc07,bra11,mic14} concentrate on mean velocities, disregarding both fluctuations and turbulent-kinetic-energy (TKE) budgets. In the present work, we aim at characterizing the flow in an idealized urban environment by running well-resolved LES of the different flow regimes identified by Oke \cite{oke88} in order to obtain one-point turbulent-flow statistics. We consider a relatively high Reynolds number and we employ tripping to assure that the flow investing the obstacles is turbulent. The analysis includes mean velocities, fluctuations, and TKE budgets. In \S\ref{sec2} we introduce the computational method and setup used during the simulations. The results of the simulations are presented and discussed in \S\ref{sec4}. Finally, we will introduce conclusions and some closing notes on other lines of investigation on urban flows, including coherent structures, in \S\ref{sec7}.

\section{Computational method and setup}\label{sec2}
The flow of air in urban environments is characterized by relatively low velocities, well below the speed of sound. Thus, the incompressible Navier--Stokes equations can be used to model the flow. These equations have been solved using the computational-fluid-dynamics (CFD) code Nek5000, which was developed by Fischer \textit{et al.}~\cite{fis08}. Nek5000 is based on the spectral-element method (SEM) developed by Patera \textit{et al.}~\cite{pat84}, which combines the geometrical flexibility of the finite-elements method (FEM) with the accuracy of the global spectral methods. Within the elements, the governing equations are discretized using a Galerkin projection in the $\mathbb{P}_{N}$--$\mathbb{P}_{N-2}$ formulation, \textit{i.e.} where the test and trial functions are obtained in the polynomial spaces $\mathbb{P}_{N}$--$\mathbb{P}_{N-2}$ of maximum order $N$ and $N-2$ for velocity and pressure respectively. Nek5000 has been extensively used for turbulent-flow simulations in complex geometries~\citep{varghese_frankel_fischer_2007,duct_ref,vin18,pipe_ref,wing_ref}, and it is thus adequate for the urban-environment cases considered here. The turbulence statistics are computed with the toolbox developed by Vinuesa \textit{et al.}~\cite{vinuesa_toolbox}.

The complexity of turbulent urban flows requires using high-fidelity methods to resolve the relevant flow structures correctly. Direct numerical simulations (DNSs) are often used in wall-bounded turbulent flows~\citep{sim09,hoy22}; however, in the case of urban environments, the presence of obstacles forbids the use of classical tools of DNS such as fast-Fourier methods~\citep{can12,llu21c}, making computational cost unaffordable. In the present work, we conduct well-resolved LES, the resolution criteria of which is close to that of a coarse DNS. The implementation of this well-resolved LES approach in Nek5000 is extensively documented by Negi \textit{et al.}~\cite{neg18}, who obtained excellent agreement with DNS statistics in turbulent wings while significantly reducing the computational cost. In these simulations, the governing equations of the system are written in dimensionless form as:
\begin{align}
& \frac{\partial U_i}{\partial t} + U_j \frac{\partial U_i}{\partial x_j} = - \frac{\partial P}{\partial x_i} + \frac{1}{Re_h} \frac{\partial^2 U_i}{\partial x_j \partial x_j} - \mathcal{H}(U_i)\\
& \frac{\partial U_i}{\partial x_i} = 0\,.
\end{align}
The instantaneous velocity field is denoted by $\bm{U}(x,y,z,t)$, where $x$, $y$, and $z$ are the streamwise, vertical and spanwise directions, respectively, and $t$ is time. The pressure is denoted by $P$. The three components of the velocity in the spatial directions are $\bm{U}=(U,V,W)$. Note that the indexes $i$ and $j$ run from $1$ to $3$, spanning through the spatial coordinates, and that the Einstein's notation of summation for repeated indexes is applied. All length quantities are normalized using the obstacle height, denoted by $h$, and the velocity scale is the free-stream value, denoted by $U_{\infty}$. The Reynolds number $Re_h=U_{\infty} h / \nu$, is based on the free-stream velocity, the obstacle height, and the kinematic viscosity. The symbol $\mathcal H(U_i)$ denotes the LES relaxation that acts as a volume force. The volume force is implemented as a high-pass filter with a given amplitude and on a subset of modes within each spectral element. Following the Reynolds decomposition, $\bm{U}$ is defined as $\bm{U} = \overline{\bm{U}} + \bm{u}$, where $\overline{\bm{U}}$ is the average in time and $\bm{u}$ is the turbulent fluctuation. The components of the Reynolds-stress tensor are thus denoted by $\overline{u_iu_j}$.

\begin{figure}
    \centering
    \includegraphics[width=\textwidth]{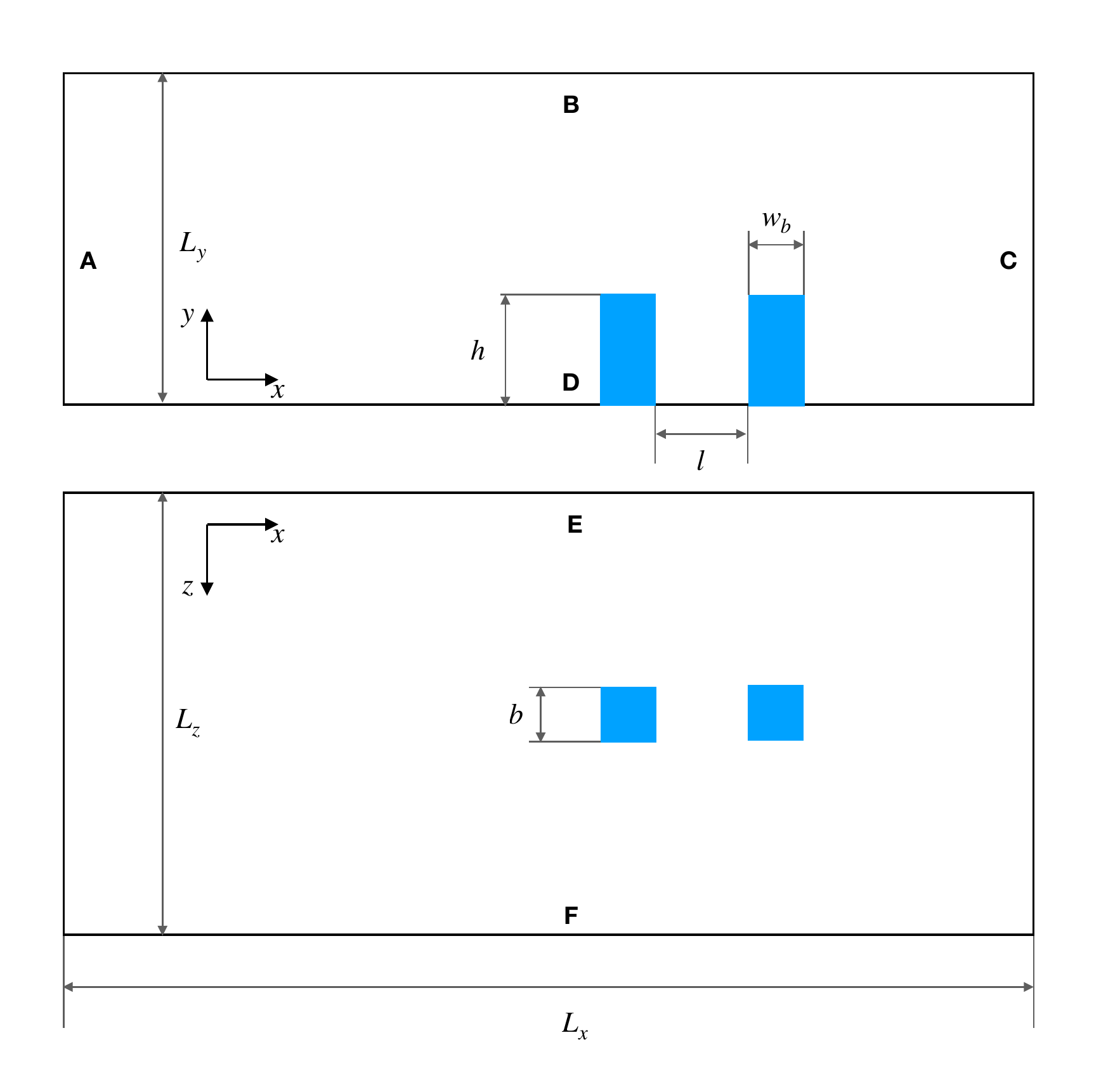}
    \caption{Schematic representation of the simulation domain, where the top and bottom panels show the side and top views, respectively. The flow goes from left to right and the obstacles are marked in blue. The center of the first obstacle is at distance of $10h$ from the inflow, and faces A, B, C, D, E, and F are the boundaries of the domain.}
    \label{fig:SchemeGeometryFinal}
\end{figure}

In Figure \ref{fig:SchemeGeometryFinal} we show a schematic representation of the geometry used in the three simulations, where $L_x$, $L_y$, and $L_z$ represent the dimensions of the computational domain in the streamwise, vertical, and spanwise directions, respectively. The vertical and spanwise dimensions are the same in the three cases, while the streamwise dimension of the domain is varied proportionally to the distance between the obstacles $l$. The obstacles are defined using three parameters: $h$, $w_b$, and $b$, their height, length, and width, respectively. Table \ref{tab:cases} gathers the geometrical data of the three cases considered in the present work. In all cases, the Reynolds number is defined with a value of $Re_h=10,000$.

As stated in the introduction, we present here the results of three different configurations, representative of the three flow regimes documented by Oke \cite{oke88}. As explained by Sini \textit{et al.}~\citep{sin96}, these configurations depend on the ratio $l/h$ as follows: if this ratio is small enough, {\it i.e.} for narrow streets, the flow above the buildings slightly penetrates on the street, in a configuration denoted by skimming flow (SF). The second situation, wake interference (WI), is present for wider streets, where the wake of the first building interacts with the second one. Finally, the third configuration is called isolated roughness (IR), and it corresponds to very broad streets, where the interaction of the wake of the first building with the second one is small. Table~\ref{tab:cases} summarizes the geometrical parameters of the three cases.

The inflow is set at face A (Figure \ref{fig:SchemeGeometryFinal}) and the outflow is set at face C. To improve our results, we have applied the stabilized outflow developed by Dong \textit{et al.}~\cite{don14}. In the spanwise direction, {\it i.e.} faces E and F, we impose periodicity. At face B we prescribe a stress-free condition in $y$, zero velocity in $z$ and we set $U_{\infty}$ in $x$. Face D and the faces that form the obstacle are set as solid walls, {\it i.e.} no-slip and no-penetration conditions.

\begin{table}
\begin{center}
    \caption{Geometrical parameters of the three flow cases under study. The reported number of grid points is based on polynomial order $N=7$, and the reported averaging periods to obtain turbulence statistics follow 40 convective time units, which are discarded to avoid initial transients. All the averaging periods correspond to over 13 eddy-turnover times, based on the $u_{\tau}$ and $h$ values of the TBL at $x/h=-2$.}
    \label{tab:cases}
    \begin{tabular}{cccccccccc}
         \textbf{Case Name}& Case code& $L_x/h$ & $L_y/h$ & $L_z/h$  & $b/h$ & $w_b/h$& $l/h$ & N. grid points & Av. period\\
         \hline
         Skimming flow &  SF &16 & 3 & 4  & 0.5 & 0.5 & 1.5 & $103 \times 10^6$ & 104 \\
         Wake interference & WI &17 & 3 & 4  & 0.5 & 0.5 & 2.5 & $116 \times 10^6$ & 104 \\
         Isolated roughness & IR & 21 & 3 & 4 & 0.5 & 0.5 & 4.5 & $142 \times 10^6$  & 105 \\
    \end{tabular}
\end{center}
\end{table}

\subsection{Turbulent-boundary-layer development}\label{subsec2.1}
As discussed in the introduction, urban flows are turbulent~\citep{oke88}. Thus, we set up the flow so that the incoming turbulent boundary layer (TBL) can develop before reaching the obstacles. In this study, the inflow condition is a Blasius laminar profile with $Re_{\delta^*}=450$, which is the Reynolds number based on displacement thickness $\delta^*$. Then, we trigger the transition to turbulence employing a numerical tripping force, acting along an horizontal line on the ground wall and at $x=-9h$. Numerical tripping is a technique that consists of introducing a weak random volume in the forcing terms of the incompressible Navier--Stokes equation acting in the wall-normal direction such that disturbances are created in the flow, as documented in Refs.~\cite{sch12,hosseini_et_al}. Next, we will discuss the turbulence statistics of the TBL upstream of the first obstacle in the SF case, noting that these results are the same in the other two cases.

\begin{figure}
\centering
    \includegraphics[width=0.99\textwidth]{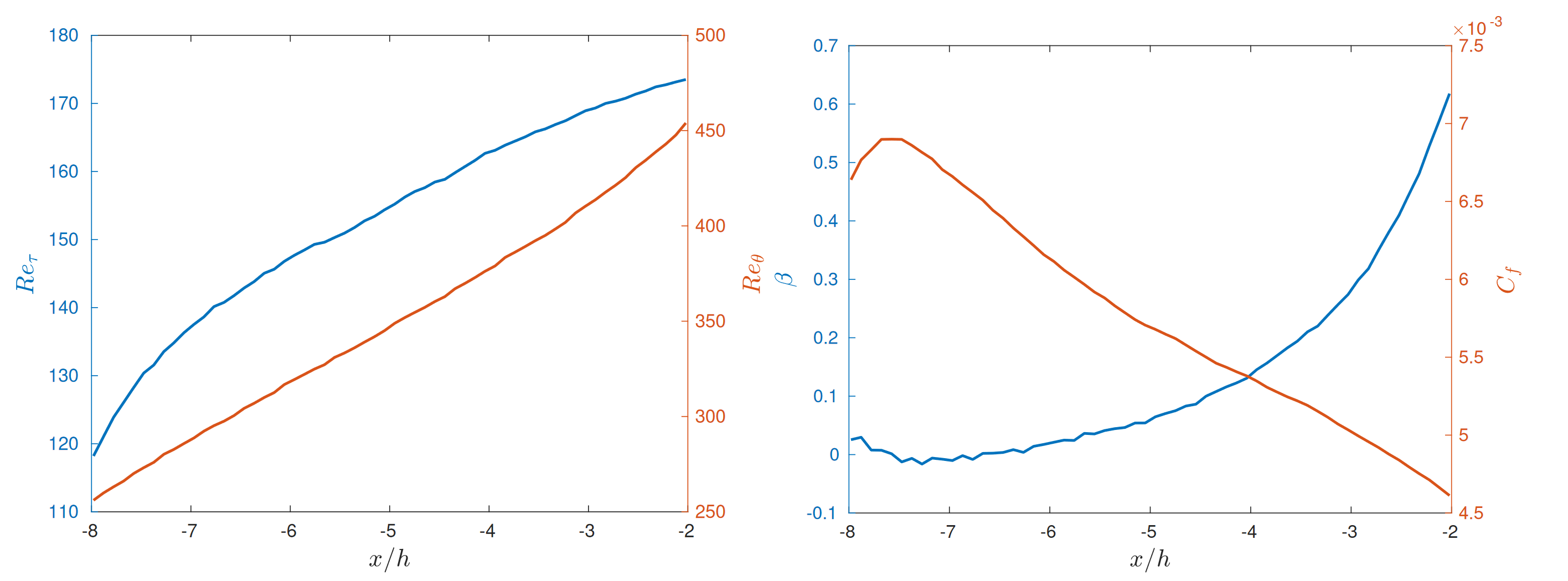}

\caption{Streamwise evolution of (left) the friction and momentum-thickness-based Reynolds numbers, and (right) the Rota--Clause parameter and the skin-friction coefficient in the region upstream of the first obstacle.}
\label{fig:BLquantities}
\end{figure}

In Figure~\ref{fig:BLquantities}~(left) we present the streamwise evolution of the  friction Reynolds number $Re_\tau=u_\tau h/\nu$ and the Reynolds number based on momentum thickness  $Re_\theta=U_\infty \theta / \nu$. Here $u_\tau=\sqrt{\tau_w/\rho}$ is the friction velocity, $\tau_w$ the wall-shear stress, $\rho$ the fluid density, $\nu$ the fluid kinematic viscosity and $\theta$ is the momentum thickness. As expected, $Re_{\theta}$ increases with the streamwise coordinate, starting at the application of the tripping force. Upstream of the first obstacle we obtain $Re_{\tau} \simeq 175$, which corresponds to fully-turbulent conditions. Note that the recirculation region upstream of the first obstacle induces an adverse pressure gradient (APG) on the TBL, which can be characterized in terms of the Rota--Clauser pressure-gradient parameter $\beta= \delta^* / \tau_w {\rm d}P_e / {\rm d}x$, where ${\rm d}P_e / {\rm d}x$ is the streamwise pressure gradient at the boundary-layer edge. This parameter, together with the skin-friction coefficient $C_f=2 (u_{\tau}/U_e)^2$ (where $U_e$ is the local edge velocity) are shown in Figure~\ref{fig:BLquantities}~(right). Note that the boundary-layer thickness is obtained using the method proposed by Vinuesa \textit{et al.}~\cite{vinuesa_pof}. The streamwise APG produces the increase of $\beta$ with $x$, reaching a value of around 0.6 at $x/h=-2$. This value corresponds to a moderate APG. The skin-friction coefficient slightly grows between $x/h=-8$ and $x/h=-7$, which is explained by the effects of the tripping force. However, $C_f$ decreases in the region upstream the first obstacle, a behavior consistent with the TBL development and the APG.

\begin{figure}
    \centering
    \includegraphics[width=0.99\textwidth]{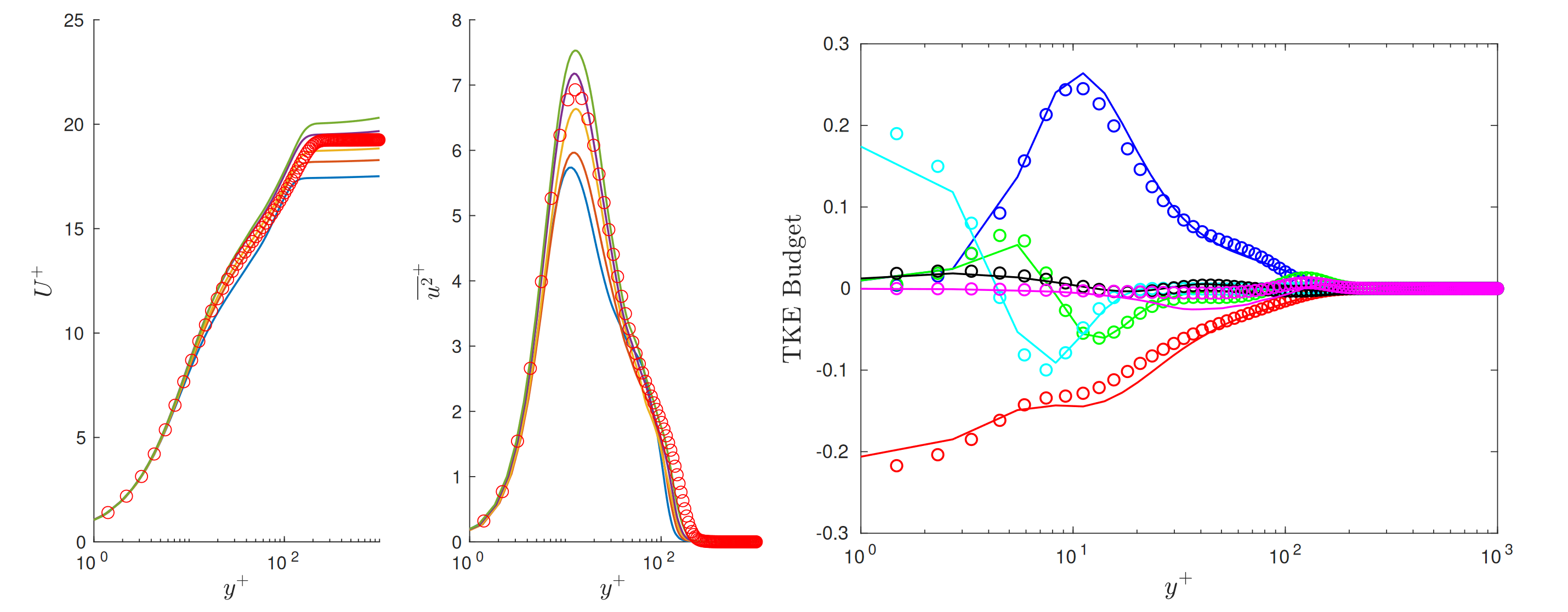}
    \caption{(Left) Inner-scaled mean and (middle) velocity-fluctuation profiles at the following streamwise locations: (blue) $x/h=-7$, (red) $x/h=-6$, (yellow) $x/h=-5$, (purple) $x/h=-4$ and (green) $x/h=-3$; the dots represent the profiles by Eitel-Amor~\cite{eit14} at the same $Re_{\tau}$ as the profile at $x/h=-4$. (Right) TKE budget terms at $x/h=-4$, where lines denote data from our simulation and dots results in Ref.~\cite{eit14} at matched $Re_{\tau}$. Blue, red, green, cyan, black and magenta represent production, dissipation, turbulent transport, viscous diffusion, velocity-pressure correlation and convection terms, respectively.}
    \label{fig:TKEBudgets_vs_ypls}
\end{figure}

The statistics presented in Figure~\ref{fig:BLquantities} are validated with the results by Eitel-Amor \textit{et al.}~\cite{eit14} for a zero-pressure-gradient (ZPG) TBL, simulated by well-resolved LES up to $Re_\theta=8,300$. In Figure~\ref{fig:TKEBudgets_vs_ypls}~(left) and (middle) we show the inner-scaled mean velocity and streamwise-velocity fluctuation profiles of our TBL at various streamwise positions, together with the profiles extracted from Ref.~\cite{eit14} at $Re_{\tau} \simeq 145$, which is the $Re_{\tau}$ of our TBL at $x/h=-4$. Note that we choose this location for comparison because here turbulence is already developed, and $\beta \simeq 0.1$, {\it i.e.} the TBL is in nearly-ZPG conditions. The various streamwise profiles reflect an adequate TBL development, and comparison at $x/h=-4$ with the ZPG TBL in Ref.~\cite{eit14} shows excellent agreement, a fact that indicates that the incoming TBL is properly simulated. Furthermore, in Figure~\ref{fig:TKEBudgets_vs_ypls}~(right), we compare the terms of the TKE budget in the incoming TBL with those of the same reference \cite{eit14}. Additional information on the calculation of all the terms can be found in the work by Vinuesa \textit{et al.}~\cite{vinuesa_toolbox}.
Interestingly, this figure shows that all the terms are in perfect agreement, including the near-wall production peak and the turbulent transport. For $y^+<3$, both the TKE dissipation and the viscous diffusion are slightly lower than the reference values, which can be attributed to the small effect of the filter in the smallest scales. Overall, the agreement is entirely satisfactory, a fact that highlights the quality of the present simulations.

\subsection{Mesh design and resolution}\label{subsec2.2}
\begin{figure}
\centering
    \includegraphics[width=0.99\textwidth]{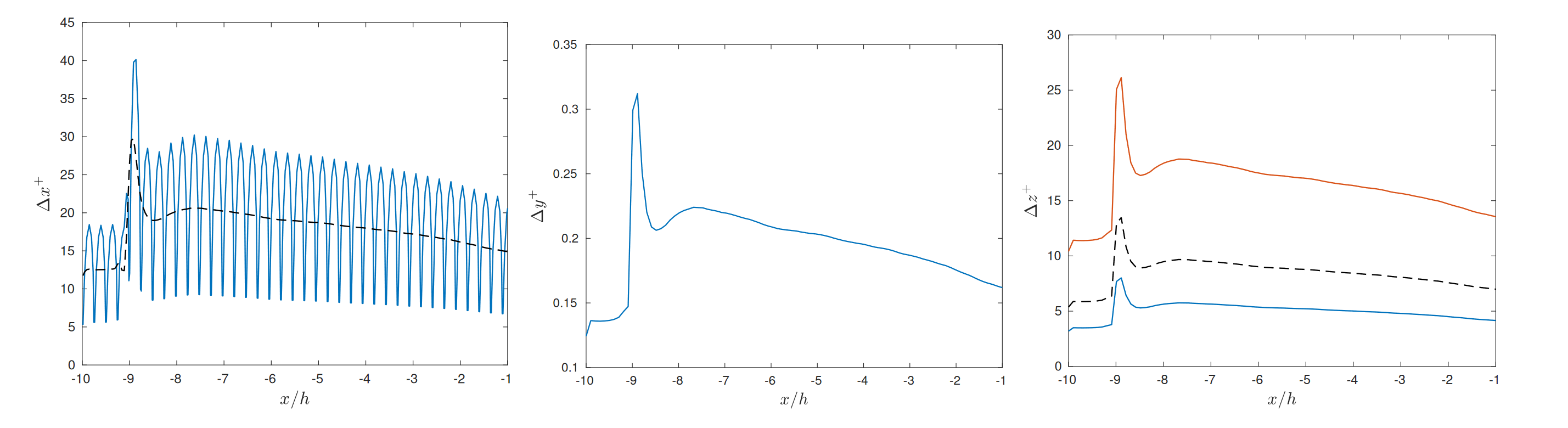}
\caption{Streamwise evolution of the inner-scaled resolution in the (left) streamwise direction, where blue denotes the local spacing and the black dashed line the average over the element. (Middle) Wall-normal resolution of the first grid point. (Right) Spanwise resolution, where blue and orange denote the minimum and maximum grid spacing of the element respectively, and the black dashed line again the average over the element.}
\label{fig:Resolution}
\end{figure}

As discussed above, Nek5000 is based on the SEM developed by Patera \cite{pat84}. The mesh comprises a number of spectral elements, ranging from 200,000 to 280,000 for the cases under consideration here, and each element has a total of $8^3$ points which follow the Gauss--Lobatto--Legendre (GLL) quadrature. The element size is refined near the wall and the obstacles in order to increase resolution. The mesh is designed following the criteria by Negi \textit{et al.}~\cite{neg18} for well-resolved LES: in the TBL part, $\Delta x^+<18$ and $\Delta z ^+<9$, which are the inner-scaled resolutions in the streamwise and spanwise directions averaged over the spectral elements. Furthermore,
$\Delta y^+<0.5$, which is the wall-normal resolution of the first grid point in inner units. In Figure~\ref{fig:Resolution} we show the streamwise evolution of these quantities for $x/h<-1$, {\it i.e.} for the region upstream of the first obstacle, and it can be observed that the criteria for well-resolved LES is satisfied within the incoming TBL. Note that the resolution in the $x$ and $z$ directions corresponds to approximately half of the one required for a DNS~\cite{hoy22}. It is interesting to note that the increase in the streamwise grid spacing at $x/h=-9$ is explained by the tripping force applied at this location. Farther from the wall, an additional requirement is satisfied for mesh resolution: defining $h=\left ( \Delta x \Delta y \Delta z\right)^{1/3}$, the ratio $h/\eta<9$ everywhere in the domain, where $\eta=(\nu^3/\varepsilon)^{1/4}$ is the Kolmogorov scale and $\varepsilon$ is the local isotropic dissipation.

\section{Results and discussion}\label{sec4}
In this section we analyze the turbulence statistics, including mean velocities, Reynolds stresses and TKE budgets in a selected portion of the computational domain. We show the statistics at the planes $y/h=0.25$ and $z/h=0$; We take advantage of the central symmetry of the case, averaging between the right and left portions of the domain for the statistics on the horizontal plane, $y/h=0.25$. Note that the following results are presented in outer scaling, {\it i.e.} in terms of $U_{\infty}$ and $h$.

\subsection{Mean flow}\label{subsec4.1}

In this section, we focus on the properties of the mean flow. The most evident effect of the increasing distance between the obstacles is the transition from a ``cavity-like'' flow and a ``wake-like'' flow in the region between the two obstacle, as already described by Zhao \textit{et al.}~\cite{zhao_et_al_pof} for a slightly different geometry and lower Reynolds number. The cavity-like flow is characterize by a very large circulation zone attached to the rear face of the first obstacle, which transports fluid in a clockwise motion. This feature of the mean flow occupies most of the space between the obstacles. The wake-like flow also exhibits a clockwise circulation zone but this is limited to the first portion of the space between the obstacles. The second portion of this space, in front of the second obstacle, is occupied by flow that is still moving in the direction of the free stream.

\begin{figure}
    \centering
    \includegraphics[width=0.99\textwidth]{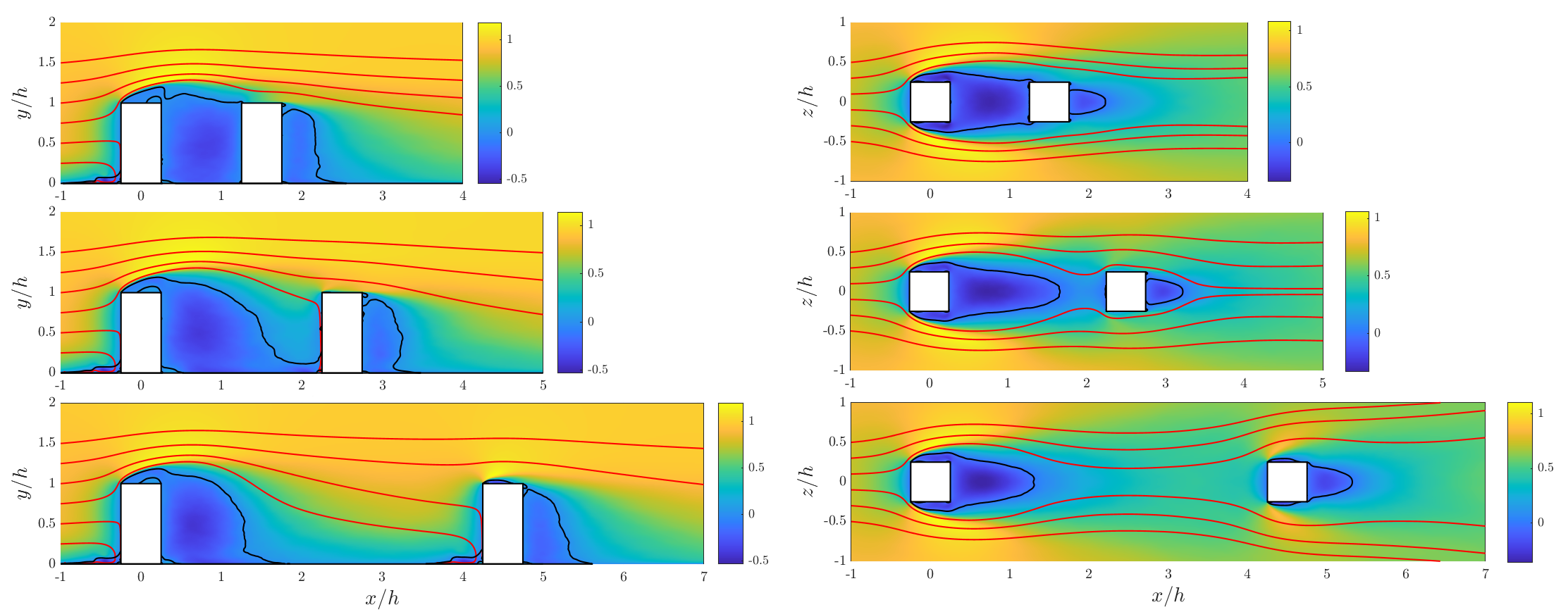}
    \caption{Mean streamwise velocity $\overline{U}$ at (left) $z/h=0$ and (right) $y/h=0.25$. The red lines and black contours denote streamlines and $\overline{U}=0$, respectively. The streamlines are computed using $\overline{U}$ and $\overline{V}$ for vertical planes and $\overline{U}$ and $\overline{W}$ for horizontal planes (note that streamlines of the 3D mean flow do not lay on the horizontal plane). From top to bottom: SF, WI and IR cases. }
    \label{fig:U}
\end{figure}
Figure~\ref{fig:U} shows the streamwise mean velocity on the planes $z/h=0$ and $y/h=0.25$, together with streamwlines computed using the mean velocity components on the two planes. In the SF case, there is only little penetration of the flow from above the canopy into the cavity. As the distance between the obstacles increases, the wake of the first obstacle becomes more apparent and there are stronger interactions between free-stream and cavity regions, as observed in the WI case. For an even higher distances between obstacles, in the IR case, the effects of the second obstacle on the wake of the first are negligible. Interestingly, the wake behind the second obstacle is relatively similar between the three regime, even thought the low speed of the incoming flow avoids the occurrence of separation at the edges in cases SF and WI.

\begin{figure}
    \centering
    \includegraphics[width=0.99\textwidth]{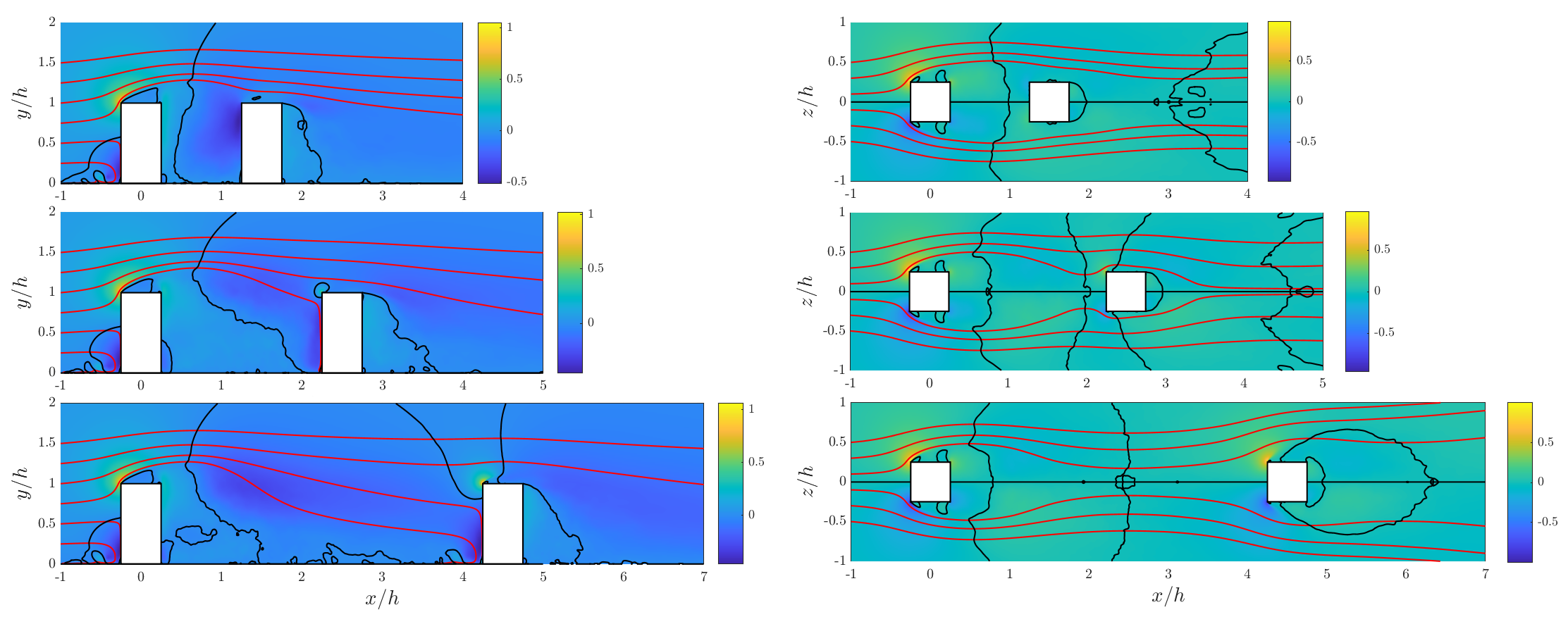}
    \caption{(Left) Mean vertical velocity $\overline{V}$ at $z/h = 0$ and (right) spanwise velocity $\overline{W}$ at $y/h = 0.25$. The red lines and black contours denote streamlines and $V=0$, respectively. The streamlines are computed using $\overline{U}$ and $\overline{V}$ for vertical planes and $\overline{U}$ and $\overline{W}$ for horizontal planes (note that streamlines of the 3D mean flow do not lay on the horizontal plane). From top to bottom: SF, WI and IR cases.}
    \label{fig:V_W}
\end{figure}

In Figure~\ref{fig:V_W} we show the vertical and spanwise mean-velocity components for the three cases. Both of these velocity components are less intense than the streamwise component in large part of the domain, with a few notable exceptions. The first exception, are the wake regions and the cavity between the obstacle in case SF, where $\overline{U}$ changes sign. The second one is the separation regions caused by the obstacle edges. These are particularly evident for the first obstacle in all three cases, and are also present in the second obstacle in the IR case. The third occurrence of where $\overline{U}$ is not the dominant mean velocity component is in regions just in front of the obstacles where the flow is deflected downwards. This is even more evident for the second obstacle than for the first one, in all cases. The effects of varying intensities and sings of the three velocity components are well summarized in the streamlines computed on the mean flow. In the WI case, where $\overline{U}$ remains high above and around the relatively short cavity, most streamlines with origin before the first obstacle pass over or to the sides of the cavity. The longer cavities in cases SF and IR however correspond to a longer region of deceleration before the second obstacle. In this region, where $\overline{V}$ is negative, streamlines laying on the vertical plane are deflected downwards.

The three flow regimes also differ in how the signs of $\overline{V}$ and $\overline{W}$ change in the domain, which is particularly affected by the change of regime between cases SF and WI. In case SF, with the large zone of clockwise mean motion, $\overline{V}$ is negative in most of the region between the obstacles, resembling the pattern of the classical two-dimensional lid-driven cavity. In cases WI and IR, the region with positive $\overline{V}$ behind the first obstacle expands. In these cases the mean flow is still moving downstream in the higher portion of the wake, but upstream in the lower one. The topology of the mean-spanwise velocity, $\overline{W}$, at intermediate heights, such as $y/h=0.25$, is particularly interesting. In front of the first obstacles, the flow is deflected sideways, around the front edges, and it also moves from the center plane towards the outside of the cavity in the wake. In the SF case, $\overline{W}$ changes sing only once, so that the flow moves toward the center of the domain in the region in front of the second obstacle. In the WI and IR cases however, at this $y/h$, $\overline{W}$ changes sing at least twice, so that the flow in the cavity moves outwards behind the first obstacle, inwards afterwards, and outwards again before the second obstacle. These differences in the topology of $\overline W$ are yet another aspect of the modification of the mean flow between case SF, where the second object is completely engulfed by the wake of the first one, and cases WI and IR, where the wake of the first object is confined between the obstacles. In case IR, $\overline W$ also shows the further development of the wake flow around the second obstacle, due to the appearance of mean separation over the side faces.

There are both similarities and differences between our results and those reported by Zhao \textit{et al.}~\cite{zhao_et_al_pof}, who also studied the flow around two obstacles but with a lower width-to-height ratio of $w_b/h=0.25$, with laminar incoming flow, and at $Re_h=500$. They considered distances between obstacles up to $l/h=2$, corresponding with the first two regimes that the examined. The transition between a cavity-like flow in the SF regime and a wake-like flow in the WI regime is also observed, but it already occurs for $l/h=1.25$, which is a distance lower than that of SF case in our dataset ($l/h=1.5$). The wake behind the second obstacle seems longer for SF configurations than for WI configurations in the database studied in Ref. \cite{zhao_et_al_pof}, a phenomenon that is not as evident in our data. This sort of comparison is however made difficult by the fact that both $Re_h$ and $w_b/h$ are different between the two studies.

\subsection{Reynolds stresses}

\begin{figure}
    \centering
    \includegraphics[width=0.99\textwidth]{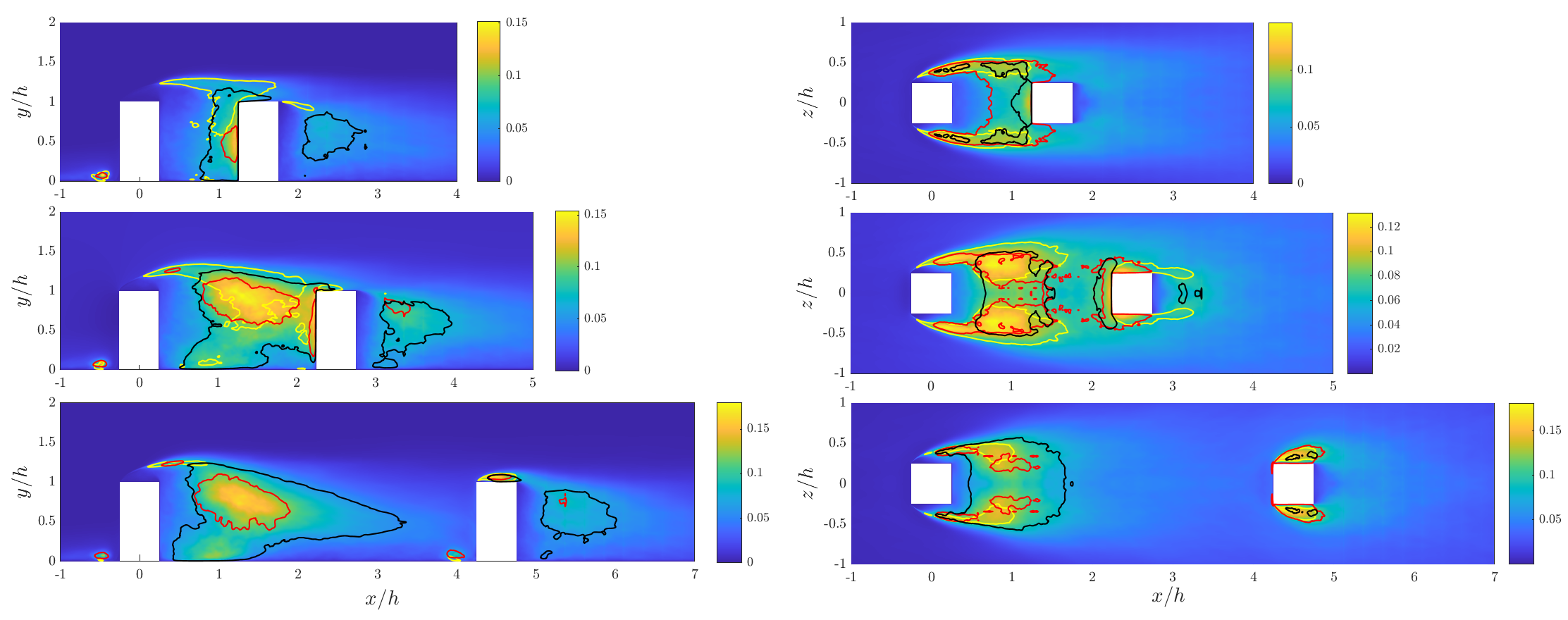}
    \caption{Turbulent kinetic energy, denoted by $k$, at (left) $z/h = 0$ and (right) $y/h = 0.25$. The yellow, red, and black contours denote regions of high $\overline{u^2}$,  $\overline{v^2}$, and  $\overline{w^2}$ respectively. In these regions, the considered quantity is higher than $1/3$ of its maximum in the domain. From top to bottom: SF, WI and IR cases.}
    \label{fig:k_uu}
\end{figure}

The Reynolds stresses show the distribution of the turbulence fluctuations within the domain.
In Figure~\ref{fig:k_uu} we illustrate the turbulent kinetic energy, denoted by $k=1/2(\overline{u^2}+\overline{v^2}+\overline{w^2})$, as well as contours highlighting regions of higher values for each of the three components of the Reynolds stress.
Turbulent fluctuations tend to be more intense between the two obstacles in all the cases and the horseshoe vortex in front of the obstacles (when present). The highest values of $k$ tend to be located after approximately $1$ unit length downstream of the first obstacle. In the SF case, where the cavity is particularly short, this region of intense fluctuations is adjacent to the front face of the second obstacle. Both the first obstacle wake and the region in front of the second one exhibit high fluctuations in the WI case. In the IR case, the region of higher $k$ first obstacle wake does not reach the second one due to an even more extended cavity. However, the first obstacle's influence on the second is still very apparent. The most intense fluctuations around the second obstacle are generated in the separation region around the obstacle edges rather than in the wake. Note that this finding seems to contrast with the description of Oke \cite{oke88}, who found only negligible effects for the flow around the downstream obstacle for a similar cavity length.

The three diagonal components of the Reynolds-stress tensor have their maxima in different positions. The highest values of the streamwise normal stress are found in the upper region of the wake, in the high-shear flow immediately following the separation bubble on top of the first obstacle and, in the IR case, in the turbulent region of separation on top of the second obstacle. As the distance between the obstacles increases, the streamwise normal-Reynolds stress values also increase in the region between the obstacle. This increment of $\overline{u^2}$ is probably connected with the interaction between low-momentum flow in the wake and high-momentum flow in the free-stream. In fact, in the WI case -- where the interaction between the layers of fluid is maximum within the cavity -- the largest region of high values of $\overline{u^2}$ is observed. In addition, the values of $\overline{u^2}$ in the second obstacle wake are larger than those of the SF case. In the IR case, we observe that the region of high $\overline{u^2}$ between the two obstacles remains attached to the first one.

Regarding the vertical normal component of the Reynolds stresses (denoted by $\overline{v^2}$), in the SF regime, we find a region of high values attached to the front wall of the second obstacle. The strong fluctuations at the front wall of the second obstacle can be explained by interactions between the high-momentum fluid moving from the free stream -- that descends into the cavity parallel to the front wall of the second obstacle -- and the low-momentum fluid in the cavity. An increase in the distance between the obstacle, as we can see for the WI case, produces a new region of high $\overline{v^2}$ at the centre of the cavity, as the miximing between flow in the cavity and flow outside the cavity becomes more pronounced. At the wake of the second obstacle, the region of high $\overline{v^2}$ is extended from that in the SF case. This extension is the result of the overall increase of turbulent fluctuations around the second obstacle, which is invested by flow with higher speed. In the IR case, intense vertical fluctuations are not present anymore in the region in front of the second obstacle.

The spanwise fluctuations, $\overline{w^2}$, reflect the same trend as the other normal components of the Reynolds stress tensor, and further confirm the significant impact that the presence of the first obstacle still has on the second in the IR case.

\begin{figure}
    \centering
    \includegraphics[width=0.99\textwidth]{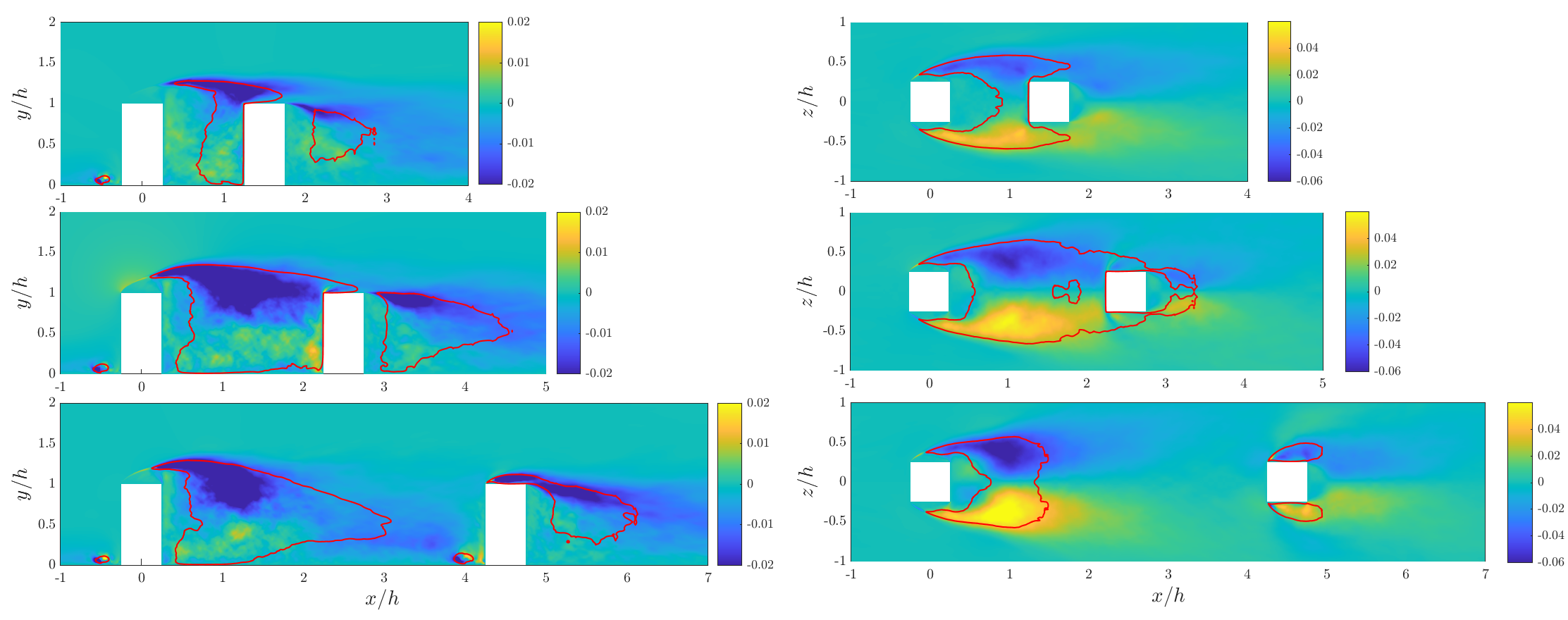}
    \caption{Contour plots of the shear-Reynolds stresses (left) $\overline{uv}$, at $z/h = 0$ and, (right) $\overline{uw}$ at $y/h = 0.25$. Note that a symmetric range of values is chosen for all figures, to help distinguish positive and negative values. The red contours denote regions of high turbulent kinetic energy, $k$, reported as reference. In these regions, $k$ is higher than $1/3$ of its maximum in the domain. From top to bottom: SF, WI and IR cases.}
    \label{fig:uv_uu}
\end{figure}

The shear-Reynolds stresses $\overline{uv}$ and $\overline{uw}$ are shown in Figure \ref{fig:uv_uu} (left and right, respectively). These quantities allow to discuss the prevalent orientation of turbulent fluctuations. They both tend to be particularly intense in regions where $\overline{u^2}$ also have higher values, \textit{i.e.} in the upper region of the wakes and, in the IR case, in the turbulent region of separation on top of the second obstacle. The vertical shear-Reynolds stress, $\overline{uv}$, is negative in the regions where it is more intense, showing that turbulent fluctuations tend in general to drive momentum downwards into the cavity. The extension of regions with different sign however varies between the three flow regimes. In the SF case, $\overline{uv}$ is positive in most of the cavity, including relatively vast portion of space where the mean vertical velocity, $\overline{V}$, is negative. In this case, mean convection contrasts turbulent transport. In the WI case, which is the case with stronger mixing between the low-speed flow in the wake and the high-speed flow outside the wake, the region of negative and intense $\overline{uv}$ occupies the higher portion of the cavity. The region of positive $\overline{uv}$ underneath includes the location with the highest positive $\overline{uv}$ observed in the three cases, which is attached to the lower portion of the front face of the second obstacle. In this region again, relatively intense turbulent fluctuations have opposite orientation than the mean convection. Lastly, in the IR case, the region of intense negative $\overline{uv}$ is also limited to approximately $2$ length units downstream the first obstacle.

The horizontal shear-Reynolds stress, $\overline{uw}$, exhibits a similar behaviour to that of $\overline{uv}$, as turbulent fluctuations carries momentum towards the inner region of cavity. This term of the Reynolds stress however tends to be of higher values then $\overline{uv}$, because fluctuations in the spanwise direction are not limited by the presence of the ground as those in the vertical direction are. In particular, the patter of $\overline{uw}$ on the left and the right of the obstacles, tends to reproduce that of $\overline{uv}$ in the upper half of the cavities, in all cases.

\subsection{Budget of the turbulent kinetic energy}
\begin{figure}
    \centering
    \includegraphics[width=0.99\textwidth]{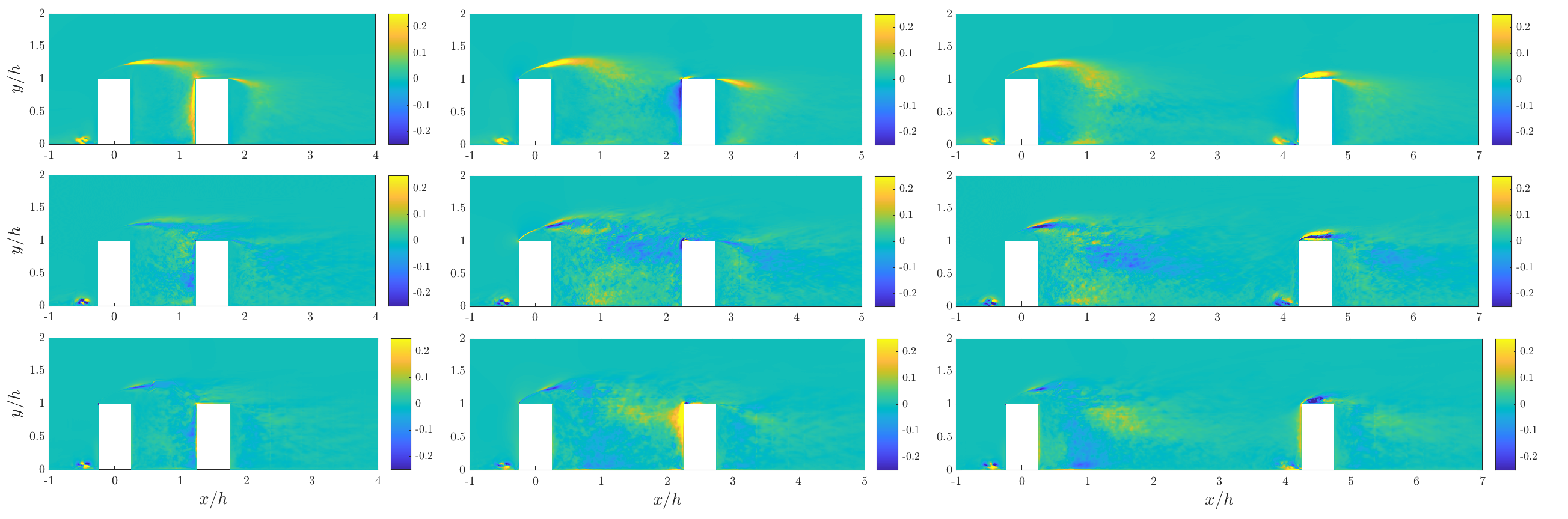}
    \caption{Selected terms of the turbulent-kinetic-energy budget on the vertical plane $z/h=0$. From top to bottom: production, turbulent diffusion, and velocity-pressure-gradient correlation, denoted by $P_k$, $T_k$, and $\Pi_k$, respectively. From left to right: SF, WI and IR cases. Note that the same symmetric color map is used, to highlight positive and negative values in all cases, even though it may not properly represent maxima and minima.}
    \label{fig:TKEBudget_vertical}
\end{figure}

\begin{figure}
    \centering
    \includegraphics[width=0.99\textwidth]{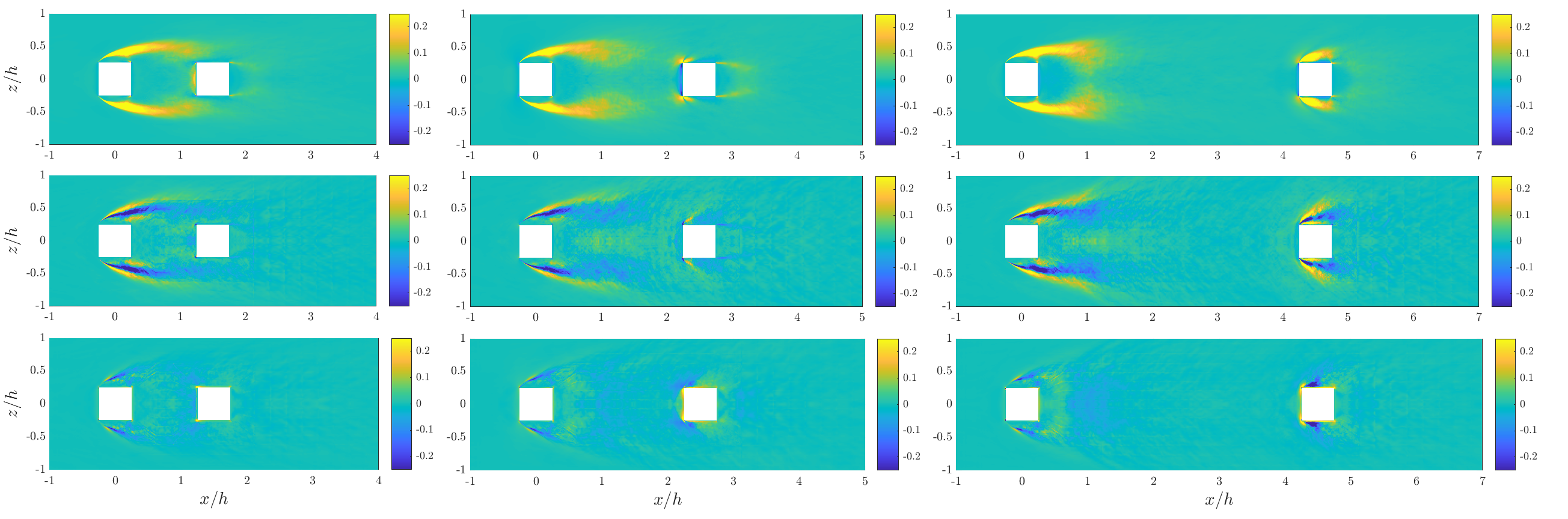}
    \caption{Selected terms of the turbulent-kinetic-energy budget on the horizontal plane $y/h=0.25$. From top to bottom: production, turbulent diffusion, and velocity-pressure-gradient correlation, denoted by $P_k$, $T_k$, and $\Pi_k$, respectively. From left to right: SF, WI and IR cases. Note that the same symmetric color map is used, to highlight positive and negative values in all cases, even though it may not properly represent maxima and minima.}
        \label{fig:TKEBudget_horizontal}
\end{figure}

To close the present discussion we will analyse terms of the turbulent-kinetic-energy (TKE) budget. We show production, turbulent diffusion, and velocity-pressure-gradient correlation in Figures~ \ref{fig:TKEBudget_vertical} and \ref{fig:TKEBudget_horizontal} for the vertical plane $z/h=0$ and the horizontal plane $y/h=0.25$, respectively. Each quantity is defined as described by Pope~ \cite{pop00}.

All terms of the TKE budget are virtually negligible in the free stream,  away from the obstacles. On the other hand, most terms exhibit relatively high values in the horseshoe vortexes. These vortexes are always present in front of the first obstacle and the second one in the IR case. In the cavity and the proximity of the obstacles, the three flow regimes differ significantly.

In the SF case, turbulent production, denoted by $P_k$, is particularly intense in three regions, \textit{i.e.} immediately downstream of the separation region on top of the first obstacle, in front of the front face of the second obstacle, and downstream the trailing edge of the second obstacle. In this case, $P_k$ is almost negligible within most of the cavity. In the WI case, a relatively large region with negative $P_k$ appears in front of the second obstacle, and production occurs within the cavity.
%\sergio{Do not totally understand this sentence} \marcoc{I used ``close'' in a funny way, I substitute it with ``just''}
In the IR case, $P_k$ in the cavity is more intense just behind the first obstacle.  Around the second obstacle, however, the most intense production occurs in the regions of separated flow adjacent to the faces. Interestingly, $P_k$ remains negative in front of the second obstacle in case IR, even though it is much less intense in that region than in case WI. The contours on the horizontal plane highlight the importance of high-shear regions on the sides faces of the first obstacles, which are also relatively similar for all cases.

Turbulent diffusion, denoted by $T_k$, tends to be negative in regions of very high turbulent fluctuations and positive in the adjacent regions where their intensity rapidly decreases. This fact is particularly apparent on the horizontal plane.
%\sergio{Remove $v^2$ definition} \marcoc{Do you mean the $\overline{v^2}$ in the next sentence, right? I intended only to say that $\overline{v^2}$ is most significant component, it was not a definition: I wrote it again to make it more clear}.
In the SF case, $T_k$ is almost negligible in most of the cavity, where fluctuations are low, and it has negative values in front of the second obstacle, where we found the highest values of $\overline{v^2}$. In both cases WI and IR, the turbulent diffusion is negative in the upper region of the wake and positive in the lower region, but, in case IR, $T_k$ becomes negligible before the second obstacle.

%\sergio{This paragraph is a bit obscure} \marcoc{Yes, it was a little convoluted and there were a couple of typos. My intention in this paragraph is to focus on behaviour of $\Pi_k$ just in front of the second obstacle, I changed it and hope is better now.}
The velocity-pressure-gradient correlation, denoted by $\Pi_k$, is negative in regions of very intense turbulent fluctuations, which is a trend similar to what observed for the turbulent diffusion, $T_k$.
The change of sign of $\Pi_k$ in the region in front of the second obstacle is however particularly interesting. In case SF, this term of the TKE budget is negative in that region. In case WI, $\Pi_k$ becomes positive and reaches relatively high values there. In case IR, $\Pi_k$ remains positive in front of the second obstacle, but its value decreases significantly. The qualitative behaviour of $\Pi_k$ is then opposite to that of the production term, in this particular region of the domain.

Regarding the other terms of the TKE budget that are not shown here, they are viscous diffusion, pseudo dissipation, and convection. Viscous diffusion is negligible, except for small near-wall regions in all cases. Pseudo dissipation is also relatively low, if compared to the other terms of the budgets in most of the domain. Convection, denoted by $C_k$, reaches it highest positive values in the high-shear regions related to separation, similarly as it happens for the other terms already described. In cases SF and WI, in the cavity, it has negative values in proximity of the front face of the second obstacle, where the mean vertical velocity is negative and carries turbulent kinetic energy downwards. In cases WI and IR, where a wake region is clearly distinguishable downstream of the first obstacle, $C_k$ exhibits positive values in the upper region of the wake and negative values in the lower region, which is qualitatively the opposite pattern than that of turbulent diffusion.

\section{Conclusions}\label{sec7}

We studied the results of highly-resolved large-eddy simulations of the flow around two rectangular obstacles invested by a turbulent boundary layer. The Reynolds number based on the free-stream velocity and obstacle height is $Re_h=10,000$, whereas the incoming boundary layer reaches approximately $Re_\tau=170$ before the first obstacle. We considered three different configurations with various distances between the obstacles, corresponding to the three different flow regimes identified in the literature.

In the first regime, denoted by skimming flow (SF), there is little penetration from the free stream into the cavity between the two obstacles, and the wake of the first one engulfs the second obstacle. In this case, the topology of the mean-velocity components is more similar to a cavity than a wake flow. In the second regime, denoted by wake interference (WI), there are strong interactions between the free stream and the wake of the first obstacle. In the third regime, denoted by isolated roughness (IR), the wake behind the first obstacle is not significantly affected by the presence of the second obstacle. In the first two cases, the reduced velocity of the incoming flow prevents separation from the edges of the second obstacle. In the IR case, separation around the second obstacle occurs, but the separation bubbles are smaller than those for the first obstacle.

The inflow conditions and the relatively high Reynolds number in our study, compared to previous numerical works with multiple obstacles, allow us to discuss in detail the turbulent flow properties in the three regimes. This discussion includes the distribution of turbulent fluctuations and terms of the turbulent-kinetic-energy (TKE) budget, which are not considered in previous studies on similar geometries. Our analysis identify three critical regions of the domain that are fundamentally affected by the increasing distance between the obstacles:

i) The first region is immediately behind the first obstacle. This region is occupied by cavity-like flow in case SF and wake-like flow in cases WI and IR. The cavity flow exhibits lower turbulent fluctuations and TKE production to the wake flow.

ii) The second region is adjacent to the second obstacle's front face. This region immediately reflects the more effective flow penetration from the free stream as the distances between obstacles increase. In the SF case, we have the most intense turbulent fluctuations and production of turbulent kinetic energy in this region. In the WI case, relatively intense turbulent fluctuations are still observed, but the production term of the TKE is negative, and turbulent kinetic energy is created with more complex mechanisms, as shown by the velocity-pressure-gradient correlation. In the IR case, turbulent fluctuations are less intense, and the classical horseshoe vortex is also present in front of the obstacle.

iii) The last region, which more dramatic changes due to the increasing obstacle distance, immediately surrounds the second obstacle. In the cases SK and WI, there is no separation. In the case IR, there is mean separation and very intense turbulent fluctuations in the separation bubbles attached to the three faces.

There are also two regions of the domain where the increasing distance between obstacles has only mild repercussions. These are the surroundings of the first obstacle, as expected, and, perhaps more surprisingly, the wake of the second obstacle.

The results that we just summarized contribute to characterizing some of the possible flow configurations in what can be considered the most elementary unity of an urban environment. Nevertheless, our study has obvious limitations if that general context is considered.

The first limitation is the still very low Reynolds number compared with realistic length and velocity scales. It is reassuring to observe that our results are qualitatively similar to those in \textit{e.g.} the study by Zhao \textit{et al.}~\cite{zhao_et_al_pof} conducted at a Reynolds number twenty times lower than the present one. Nevertheless, our results also show the complexity of the turbulent flow in the three regimes, confirming that numerical studies at even higher Reynolds numbers may be required.

The second limitation is the simplicity of our configuration, which is evident in both inflow conditions and the obstacle geometry.
The crucial phenomenon in the flow that originates from a boundary layer impacting a group of obstacles is the interaction of the wake created by leading obstacles with subsequent ones. In our idealized study case, this phenomenon was governed by the only geometrical parameter that we let vary, \textit{i.e.} the obstacle distance. However, different obstacle alignments with the inflow velocity as well as different aspect ratios and relative sizes will lead to an even greater variety of flow regimes, indicating other possible directions for future investigations.

A further extension of the present study worth considering pertains to additional methodologies to characterize the different flow regimes. One approach is to examine coherent structures and link features of the instantaneous flow with the turbulence statistics that we considered here. A second approach is to determine how the three flow regimes differ in the dispersion of passive scalars and particles with inertia, which are both relevant in applications connected with pollution and pathogen contamination.

\begin{acknowledgments}
RV acknowledges the financial support of the G\"oran Gustafsson foundation. SLC acknowledges the grant PID2020-114173RB-I00 funded by MCIN/AEI/10.13039/501100011033. MA acknowledges financial support of the Austrian Science Fund (FWF), project number: I5180-N. This work was also supported by RTI2018-102256-B-I00 of MINECO/FEDER. The meshes were generated using the SciDAX platform, which was developed by Parallel Works Inc. with support from the US Department of Energy, Office of Science, under contract DE-SC0019695. The computations carried out in this study were made possible by resources provided by the Swedish National Infrastructure for Computing (SNIC).
\end{acknowledgments}

\bibliography{main}% Produces the bibliography via BibTeX.

\begin{thebibliography}{46}%
\makeatletter
\providecommand \@ifxundefined [1]{%
 \@ifx{#1\undefined}
}%
\providecommand \@ifnum [1]{%
 \ifnum #1\expandafter \@firstoftwo
 \else \expandafter \@secondoftwo
 \fi
}%
\providecommand \@ifx [1]{%
 \ifx #1\expandafter \@firstoftwo
 \else \expandafter \@secondoftwo
 \fi
}%
\providecommand \natexlab [1]{#1}%
\providecommand \enquote  [1]{``#1''}%
\providecommand \bibnamefont  [1]{#1}%
\providecommand \bibfnamefont [1]{#1}%
\providecommand \citenamefont [1]{#1}%
\providecommand \href@noop [0]{\@secondoftwo}%
\providecommand \href [0]{\begingroup \@sanitize@url \@href}%
\providecommand \@href[1]{\@@startlink{#1}\@@href}%
\providecommand \@@href[1]{\endgroup#1\@@endlink}%
\providecommand \@sanitize@url [0]{\catcode `\\12\catcode `\$12\catcode
  `\&12\catcode `\#12\catcode `\^12\catcode `\_12\catcode `\%12\relax}%
\providecommand \@@startlink[1]{}%
\providecommand \@@endlink[0]{}%
\providecommand \url  [0]{\begingroup\@sanitize@url \@url }%
\providecommand \@url [1]{\endgroup\@href {#1}{\urlprefix }}%
\providecommand \urlprefix  [0]{URL }%
\providecommand \Eprint [0]{\href }%
\providecommand \doibase [0]{https://doi.org/}%
\providecommand \selectlanguage [0]{\@gobble}%
\providecommand \bibinfo  [0]{\@secondoftwo}%
\providecommand \bibfield  [0]{\@secondoftwo}%
\providecommand \translation [1]{[#1]}%
\providecommand \BibitemOpen [0]{}%
\providecommand \bibitemStop [0]{}%
\providecommand \bibitemNoStop [0]{.\EOS\space}%
\providecommand \EOS [0]{\spacefactor3000\relax}%
\providecommand \BibitemShut  [1]{\csname bibitem#1\endcsname}%
\let\auto@bib@innerbib\@empty
%</preamble>
\bibitem [{\citenamefont {{United Nations}}(2020)}]{UNO20}%
  \BibitemOpen
  \bibfield  {author} {\bibinfo {author} {\bibnamefont {{United Nations}}},\
  }\href@noop {} {\emph {\bibinfo {title} {Cities and Pollution Contribute to
  Climate Change}}}\ (\bibinfo {year} {2020})\ \bibinfo {note}
  {\url{https://www.un.org/en/climatechange/cities-pollution.shtml, Accessed on
  10 September 2020}}\BibitemShut {NoStop}%
\bibitem [{\citenamefont {{European Environment Agency}}(2019)}]{EEA19}%
  \BibitemOpen
  \bibfield  {author} {\bibinfo {author} {\bibnamefont {{European Environment
  Agency}}},\ }\href@noop {} {\emph {\bibinfo {title} {Air Quality in Europe:
  2019 Report}}}\ (\bibinfo  {publisher} {Publications Office of the European
  Union},\ \bibinfo {year} {2019})\BibitemShut {NoStop}%
\bibitem [{\citenamefont {Lelieveld}\ \emph {et~al.}(2019)\citenamefont
  {Lelieveld}, \citenamefont {Klingm\"uller}, \citenamefont {Pozzer},
  \citenamefont {P\"oschl}, \citenamefont {Fnais}, \citenamefont {Daiber},\
  and\ \citenamefont {M\"unzel}}]{lelieveld_et_al}%
  \BibitemOpen
  \bibfield  {author} {\bibinfo {author} {\bibfnamefont {J.}~\bibnamefont
  {Lelieveld}}, \bibinfo {author} {\bibfnamefont {K.}~\bibnamefont
  {Klingm\"uller}}, \bibinfo {author} {\bibfnamefont {A.}~\bibnamefont
  {Pozzer}}, \bibinfo {author} {\bibfnamefont {U.}~\bibnamefont {P\"oschl}},
  \bibinfo {author} {\bibfnamefont {M.}~\bibnamefont {Fnais}}, \bibinfo
  {author} {\bibfnamefont {A.}~\bibnamefont {Daiber}},\ and\ \bibinfo {author}
  {\bibfnamefont {T.}~\bibnamefont {M\"unzel}},\ }\bibfield  {title} {\bibinfo
  {title} {{Cardiovascular disease burden from ambient air pollution in Europe
  reassessed using novel hazard ratio functions}},\ }\href@noop {} {\bibfield
  {journal} {\bibinfo  {journal} {Eur. Heart J.}\ }\textbf {\bibinfo {volume}
  {40}},\ \bibinfo {pages} {1590} (\bibinfo {year} {2019})}\BibitemShut
  {NoStop}%
\bibitem [{\citenamefont {Isyumov}(1978)}]{isy78}%
  \BibitemOpen
  \bibfield  {author} {\bibinfo {author} {\bibfnamefont {N.}~\bibnamefont
  {Isyumov}},\ }\bibfield  {title} {\bibinfo {title} {Studies of the pedestrian
  level wind environment at the boundary layer wind tunnel laboratory of the
  university of western ontario},\ }\href
  {https://doi.org/https://doi.org/10.1016/0167-6105(78)90009-0} {\bibfield
  {journal} {\bibinfo  {journal} {Journal of Wind Engineering and Industrial
  Aerodynamics}\ }\textbf {\bibinfo {volume} {3}},\ \bibinfo {pages} {187}
  (\bibinfo {year} {1978})},\ \bibinfo {note} {the wind content of the built
  environment}\BibitemShut {NoStop}%
\bibitem [{\citenamefont {Britter}\ and\ \citenamefont {Hunt}(1979)}]{bri79}%
  \BibitemOpen
  \bibfield  {author} {\bibinfo {author} {\bibfnamefont {R.}~\bibnamefont
  {Britter}}\ and\ \bibinfo {author} {\bibfnamefont {J.}~\bibnamefont {Hunt}},\
  }\bibfield  {title} {\bibinfo {title} {Velocity measurements and order of
  magnitude estimates of the flow between two buildings in a simulated
  atmospheric boundary layer},\ }\href
  {https://doi.org/https://doi.org/10.1016/0167-6105(79)90044-8} {\bibfield
  {journal} {\bibinfo  {journal} {Journal of Wind Engineering and Industrial
  Aerodynamics}\ }\textbf {\bibinfo {volume} {4}},\ \bibinfo {pages} {165}
  (\bibinfo {year} {1979})}\BibitemShut {NoStop}%
\bibitem [{\citenamefont {Zajic}\ \emph {et~al.}(2011)\citenamefont {Zajic},
  \citenamefont {Fernando}, \citenamefont {Calhoun}, \citenamefont {Princevac},
  \citenamefont {Brown},\ and\ \citenamefont {Pardyjak}}]{zaj11}%
  \BibitemOpen
  \bibfield  {author} {\bibinfo {author} {\bibfnamefont {D.}~\bibnamefont
  {Zajic}}, \bibinfo {author} {\bibfnamefont {H.~J.~S.}\ \bibnamefont
  {Fernando}}, \bibinfo {author} {\bibfnamefont {R.}~\bibnamefont {Calhoun}},
  \bibinfo {author} {\bibfnamefont {M.}~\bibnamefont {Princevac}}, \bibinfo
  {author} {\bibfnamefont {M.~J.}\ \bibnamefont {Brown}},\ and\ \bibinfo
  {author} {\bibfnamefont {E.~R.}\ \bibnamefont {Pardyjak}},\ }\bibfield
  {title} {\bibinfo {title} {Flow and turbulence in an urban canyon},\ }\href
  {https://doi.org/10.1175/2010JAMC2525.1} {\bibfield  {journal} {\bibinfo
  {journal} {Journal of Applied Meteorology and Climatology}\ }\textbf
  {\bibinfo {volume} {50}},\ \bibinfo {pages} {203 } (\bibinfo {year}
  {2011})}\BibitemShut {NoStop}%
\bibitem [{\citenamefont {Torres}\ \emph {et~al.}(2021)\citenamefont {Torres},
  \citenamefont {Le~Clainche},\ and\ \citenamefont {Vinuesa}}]{tor21}%
  \BibitemOpen
  \bibfield  {author} {\bibinfo {author} {\bibfnamefont {P.}~\bibnamefont
  {Torres}}, \bibinfo {author} {\bibfnamefont {S.}~\bibnamefont
  {Le~Clainche}},\ and\ \bibinfo {author} {\bibfnamefont {R.}~\bibnamefont
  {Vinuesa}},\ }\bibfield  {title} {\bibinfo {title} {On the experimental,
  numerical and data-driven methods to study urban flows},\ }\bibfield
  {journal} {\bibinfo  {journal} {Energies}\ }\textbf {\bibinfo {volume}
  {14}},\ \href {https://doi.org/10.3390/en14051310} {10.3390/en14051310}
  (\bibinfo {year} {2021})\BibitemShut {NoStop}%
\bibitem [{\citenamefont {Oke}(1988)}]{oke88}%
  \BibitemOpen
  \bibfield  {author} {\bibinfo {author} {\bibfnamefont {T.}~\bibnamefont
  {Oke}},\ }\bibfield  {title} {\bibinfo {title} {Street design and urban
  canopy layer climate},\ }\href
  {https://doi.org/https://doi.org/10.1016/0378-7788(88)90026-6} {\bibfield
  {journal} {\bibinfo  {journal} {Energy and Buildings}\ }\textbf {\bibinfo
  {volume} {11}},\ \bibinfo {pages} {103} (\bibinfo {year} {1988})}\BibitemShut
  {NoStop}%
\bibitem [{\citenamefont {Britter}\ and\ \citenamefont {Hanna}(2003)}]{bri03}%
  \BibitemOpen
  \bibfield  {author} {\bibinfo {author} {\bibfnamefont {R.~E.}\ \bibnamefont
  {Britter}}\ and\ \bibinfo {author} {\bibfnamefont {S.~R.}\ \bibnamefont
  {Hanna}},\ }\bibfield  {title} {\bibinfo {title} {{Flow and dispersion in
  urban areas}},\ }\href@noop {} {\bibfield  {journal} {\bibinfo  {journal}
  {Annu. Rev. Fluid Mech.}\ }\textbf {\bibinfo {volume} {35}},\ \bibinfo
  {pages} {469} (\bibinfo {year} {2003})}\BibitemShut {NoStop}%
\bibitem [{\citenamefont {Di~Sabatino}\ \emph {et~al.}(2009)\citenamefont
  {Di~Sabatino}, \citenamefont {Leo}, \citenamefont {Hedquist}, \citenamefont
  {Carter},\ and\ \citenamefont {Fernando}}]{sab09}%
  \BibitemOpen
  \bibfield  {author} {\bibinfo {author} {\bibfnamefont {S.}~\bibnamefont
  {Di~Sabatino}}, \bibinfo {author} {\bibfnamefont {L.~S.}\ \bibnamefont
  {Leo}}, \bibinfo {author} {\bibfnamefont {B.~C.}\ \bibnamefont {Hedquist}},
  \bibinfo {author} {\bibfnamefont {W.}~\bibnamefont {Carter}},\ and\ \bibinfo
  {author} {\bibfnamefont {H.~J.~S.}\ \bibnamefont {Fernando}},\ }\bibfield
  {title} {\bibinfo {title} {{Results from the Phoenix Urban Heat Island (UHI)
  experiment: effects at the local, neighbourhood and urban scales}},\
  }\href@noop {} {\bibfield  {journal} {\bibinfo  {journal} {Eighth Symposium
  on the Urban Environment, Phoenix, Arizona (USA)}\ } (\bibinfo {year}
  {2009})}\BibitemShut {NoStop}%
\bibitem [{\citenamefont {Weerasuriya}\ \emph {et~al.}(2018)\citenamefont
  {Weerasuriya}, \citenamefont {Tse}, \citenamefont {Zhang},\ and\
  \citenamefont {Li}}]{wee18}%
  \BibitemOpen
  \bibfield  {author} {\bibinfo {author} {\bibfnamefont {A.~U.}\ \bibnamefont
  {Weerasuriya}}, \bibinfo {author} {\bibfnamefont {K.~T.}\ \bibnamefont
  {Tse}}, \bibinfo {author} {\bibfnamefont {X.}~\bibnamefont {Zhang}},\ and\
  \bibinfo {author} {\bibfnamefont {S.~W.}\ \bibnamefont {Li}},\ }\bibfield
  {title} {\bibinfo {title} {{A wind tunnel study of effects of twisted wind
  flows on the pedestrian-level wind field in an urban environment}},\
  }\href@noop {} {\bibfield  {journal} {\bibinfo  {journal} {Build. Environ.}\
  }\textbf {\bibinfo {volume} {128}},\ \bibinfo {pages} {225} (\bibinfo {year}
  {2018})}\BibitemShut {NoStop}%
\bibitem [{\citenamefont {Corke}\ \emph {et~al.}(1979)\citenamefont {Corke},
  \citenamefont {Nagib},\ and\ \citenamefont {Tan-Atichat}}]{corke_et_al}%
  \BibitemOpen
  \bibfield  {author} {\bibinfo {author} {\bibfnamefont {T.~C.}\ \bibnamefont
  {Corke}}, \bibinfo {author} {\bibfnamefont {H.~M.}\ \bibnamefont {Nagib}},\
  and\ \bibinfo {author} {\bibfnamefont {J.}~\bibnamefont {Tan-Atichat}},\
  }\bibfield  {title} {\bibinfo {title} {Flow near a building model in a family
  of surface layers},\ }\href@noop {} {\bibfield  {journal} {\bibinfo
  {journal} {J. Wind Eng. Ind. Aerodyn.}\ }\textbf {\bibinfo {volume} {5}},\
  \bibinfo {pages} {139} (\bibinfo {year} {1979})}\BibitemShut {NoStop}%
\bibitem [{\citenamefont {Nagib}\ and\ \citenamefont
  {Corke}(1984)}]{nagib_corke}%
  \BibitemOpen
  \bibfield  {author} {\bibinfo {author} {\bibfnamefont {H.~M.}\ \bibnamefont
  {Nagib}}\ and\ \bibinfo {author} {\bibfnamefont {T.~C.}\ \bibnamefont
  {Corke}},\ }\bibfield  {title} {\bibinfo {title} {Wind microclimate around
  buildings: characteristics and control},\ }\href@noop {} {\bibfield
  {journal} {\bibinfo  {journal} {J. Wind Eng. Ind. Aerodyn.}\ }\textbf
  {\bibinfo {volume} {16}},\ \bibinfo {pages} {1} (\bibinfo {year}
  {1984})}\BibitemShut {NoStop}%
\bibitem [{\citenamefont {Monnier}\ \emph {et~al.}(2010)\citenamefont
  {Monnier}, \citenamefont {Neiswander},\ and\ \citenamefont
  {Wark}}]{monnier_et_al}%
  \BibitemOpen
  \bibfield  {author} {\bibinfo {author} {\bibfnamefont {B.}~\bibnamefont
  {Monnier}}, \bibinfo {author} {\bibfnamefont {B.}~\bibnamefont
  {Neiswander}},\ and\ \bibinfo {author} {\bibfnamefont {C.}~\bibnamefont
  {Wark}},\ }\bibfield  {title} {\bibinfo {title} {Stereoscopic particle image
  velocimetry measurements in an urban-type boundary layer: insight into flow
  regimes and incidence angle effect},\ }\href@noop {} {\bibfield  {journal}
  {\bibinfo  {journal} {Boundary-Layer Meteorol.}\ }\textbf {\bibinfo {volume}
  {135}},\ \bibinfo {pages} {243} (\bibinfo {year} {2010})}\BibitemShut
  {NoStop}%
\bibitem [{\citenamefont {Monnier}\ \emph {et~al.}(2018)\citenamefont
  {Monnier}, \citenamefont {Goudarzi}, \citenamefont {Vinuesa},\ and\
  \citenamefont {Wark}}]{monnier_et_al2}%
  \BibitemOpen
  \bibfield  {author} {\bibinfo {author} {\bibfnamefont {B.}~\bibnamefont
  {Monnier}}, \bibinfo {author} {\bibfnamefont {S.~A.}\ \bibnamefont
  {Goudarzi}}, \bibinfo {author} {\bibfnamefont {R.}~\bibnamefont {Vinuesa}},\
  and\ \bibinfo {author} {\bibfnamefont {C.}~\bibnamefont {Wark}},\ }\bibfield
  {title} {\bibinfo {title} {Turbulent structure of a simplified urban fluid
  flow studied through stereoscopic particle image velocimetry},\ }\href@noop
  {} {\bibfield  {journal} {\bibinfo  {journal} {Boundary-Layer Meteorol.}\
  }\textbf {\bibinfo {volume} {166}},\ \bibinfo {pages} {239} (\bibinfo {year}
  {2018})}\BibitemShut {NoStop}%
\bibitem [{\citenamefont {Fernando}\ \emph {et~al.}(2010)\citenamefont
  {Fernando}, \citenamefont {Zajic}, \citenamefont {Di~Sabatino}, \citenamefont
  {Dimitrova}, \citenamefont {Hedquist},\ and\ \citenamefont
  {Dallman}}]{fer10}%
  \BibitemOpen
  \bibfield  {author} {\bibinfo {author} {\bibfnamefont {H.~J.~S.}\
  \bibnamefont {Fernando}}, \bibinfo {author} {\bibfnamefont {D.}~\bibnamefont
  {Zajic}}, \bibinfo {author} {\bibfnamefont {S.}~\bibnamefont {Di~Sabatino}},
  \bibinfo {author} {\bibfnamefont {R.}~\bibnamefont {Dimitrova}}, \bibinfo
  {author} {\bibfnamefont {B.}~\bibnamefont {Hedquist}},\ and\ \bibinfo
  {author} {\bibfnamefont {A.}~\bibnamefont {Dallman}},\ }\bibfield  {title}
  {\bibinfo {title} {Flow, turbulence, and pollutant dispersion in urban
  atmospheres},\ }\href {https://doi.org/10.1063/1.3407662} {\bibfield
  {journal} {\bibinfo  {journal} {Physics of Fluids}\ }\textbf {\bibinfo
  {volume} {22}},\ \bibinfo {pages} {051301} (\bibinfo {year}
  {2010})}\BibitemShut {NoStop}%
\bibitem [{\citenamefont {Vita}\ \emph {et~al.}(2020)\citenamefont {Vita},
  \citenamefont {Shu}, \citenamefont {Jesson}, \citenamefont {Quinn},
  \citenamefont {Hemida}, \citenamefont {Sterling},\ and\ \citenamefont
  {Baker}}]{vit20}%
  \BibitemOpen
  \bibfield  {author} {\bibinfo {author} {\bibfnamefont {G.}~\bibnamefont
  {Vita}}, \bibinfo {author} {\bibfnamefont {Z.}~\bibnamefont {Shu}}, \bibinfo
  {author} {\bibfnamefont {M.}~\bibnamefont {Jesson}}, \bibinfo {author}
  {\bibfnamefont {A.}~\bibnamefont {Quinn}}, \bibinfo {author} {\bibfnamefont
  {H.}~\bibnamefont {Hemida}}, \bibinfo {author} {\bibfnamefont
  {M.}~\bibnamefont {Sterling}},\ and\ \bibinfo {author} {\bibfnamefont
  {C.}~\bibnamefont {Baker}},\ }\bibfield  {title} {\bibinfo {title} {On the
  assessment of pedestrian distress in urban winds},\ }\href
  {https://doi.org/https://doi.org/10.1016/j.jweia.2020.104200} {\bibfield
  {journal} {\bibinfo  {journal} {Journal of Wind Engineering and Industrial
  Aerodynamics}\ }\textbf {\bibinfo {volume} {203}},\ \bibinfo {pages} {104200}
  (\bibinfo {year} {2020})}\BibitemShut {NoStop}%
\bibitem [{\citenamefont {Belcher}(2005)}]{bel05}%
  \BibitemOpen
  \bibfield  {author} {\bibinfo {author} {\bibfnamefont {S.~E.}\ \bibnamefont
  {Belcher}},\ }\bibfield  {title} {\bibinfo {title} {Mixing and transport in
  urban areas},\ }\href {https://doi.org/10.1098/rsta.2005.1673} {\bibfield
  {journal} {\bibinfo  {journal} {Philosophical Transactions of the Royal
  Society A: Mathematical, Physical and Engineering Sciences}\ }\textbf
  {\bibinfo {volume} {363}},\ \bibinfo {pages} {2947} (\bibinfo {year}
  {2005})},\ \Eprint
  {https://arxiv.org/abs/https://royalsocietypublishing.org/doi/pdf/10.1098/rsta.2005.1673}
  {https://royalsocietypublishing.org/doi/pdf/10.1098/rsta.2005.1673}
  \BibitemShut {NoStop}%
\bibitem [{\citenamefont {{Branford}}\ \emph {et~al.}(2011)\citenamefont
  {{Branford}}, \citenamefont {{Coceal}}, \citenamefont {{Thomas}},\ and\
  \citenamefont {{Belcher}}}]{bra11}%
  \BibitemOpen
  \bibfield  {author} {\bibinfo {author} {\bibfnamefont {S.}~\bibnamefont
  {{Branford}}}, \bibinfo {author} {\bibfnamefont {O.}~\bibnamefont
  {{Coceal}}}, \bibinfo {author} {\bibfnamefont {T.~G.}\ \bibnamefont
  {{Thomas}}},\ and\ \bibinfo {author} {\bibfnamefont {S.~E.}\ \bibnamefont
  {{Belcher}}},\ }\bibfield  {title} {\bibinfo {title} {{Dispersion of a
  Point-Source Release of a Passive Scalar Through an Urban-Like Array for
  Different Wind Directions}},\ }\href
  {https://doi.org/10.1007/s10546-011-9589-1} {\bibfield  {journal} {\bibinfo
  {journal} {Boundary-Layer Meteorology}\ }\textbf {\bibinfo {volume} {139}},\
  \bibinfo {pages} {367} (\bibinfo {year} {2011})}\BibitemShut {NoStop}%
\bibitem [{\citenamefont {Nakayama}\ \emph {et~al.}(2011)\citenamefont
  {Nakayama}, \citenamefont {Takemi},\ and\ \citenamefont {Nagai}}]{nak11}%
  \BibitemOpen
  \bibfield  {author} {\bibinfo {author} {\bibfnamefont {H.}~\bibnamefont
  {Nakayama}}, \bibinfo {author} {\bibfnamefont {T.}~\bibnamefont {Takemi}},\
  and\ \bibinfo {author} {\bibfnamefont {H.}~\bibnamefont {Nagai}},\ }\bibfield
   {title} {\bibinfo {title} {Les analysis of the aerodynamic surface
  properties for turbulent flows over building arrays with various
  geometries},\ }\href {https://doi.org/10.1175/2011JAMC2567.1} {\bibfield
  {journal} {\bibinfo  {journal} {Journal of Applied Meteorology and
  Climatology}\ }\textbf {\bibinfo {volume} {50}},\ \bibinfo {pages} {1692 }
  (\bibinfo {year} {2011})}\BibitemShut {NoStop}%
\bibitem [{\citenamefont {Nakayama}\ \emph {et~al.}(2012)\citenamefont
  {Nakayama}, \citenamefont {Takemi},\ and\ \citenamefont {Nagai}}]{nak12}%
  \BibitemOpen
  \bibfield  {author} {\bibinfo {author} {\bibfnamefont {H.}~\bibnamefont
  {Nakayama}}, \bibinfo {author} {\bibfnamefont {T.}~\bibnamefont {Takemi}},\
  and\ \bibinfo {author} {\bibfnamefont {H.}~\bibnamefont {Nagai}},\ }\bibfield
   {title} {\bibinfo {title} {Large-eddy simulation of urban boundary-layer
  flows by generating turbulent inflows from mesoscale meteorological
  simulations},\ }\href {https://doi.org/https://doi.org/10.1002/asl.377}
  {\bibfield  {journal} {\bibinfo  {journal} {Atmospheric Science Letters}\
  }\textbf {\bibinfo {volume} {13}},\ \bibinfo {pages} {180} (\bibinfo {year}
  {2012})},\ \Eprint
  {https://arxiv.org/abs/https://rmets.onlinelibrary.wiley.com/doi/pdf/10.1002/asl.377}
  {https://rmets.onlinelibrary.wiley.com/doi/pdf/10.1002/asl.377} \BibitemShut
  {NoStop}%
\bibitem [{\citenamefont {Coceal}\ \emph {et~al.}(2007)\citenamefont {Coceal},
  \citenamefont {DOBRE}, \citenamefont {THOMAS},\ and\ \citenamefont
  {BELCHER}}]{coc07}%
  \BibitemOpen
  \bibfield  {author} {\bibinfo {author} {\bibfnamefont {O.}~\bibnamefont
  {Coceal}}, \bibinfo {author} {\bibfnamefont {A.}~\bibnamefont {DOBRE}},
  \bibinfo {author} {\bibfnamefont {T.~G.}\ \bibnamefont {THOMAS}},\ and\
  \bibinfo {author} {\bibfnamefont {S.~E.}\ \bibnamefont {BELCHER}},\
  }\bibfield  {title} {\bibinfo {title} {Structure of turbulent flow over
  regular arrays of cubical roughness},\ }\href
  {https://doi.org/10.1017/S002211200700794X} {\bibfield  {journal} {\bibinfo
  {journal} {Journal of Fluid Mechanics}\ }\textbf {\bibinfo {volume} {589}},\
  \bibinfo {pages} {375–409} (\bibinfo {year} {2007})}\BibitemShut {NoStop}%
\bibitem [{\citenamefont {Vinuesa}\ \emph {et~al.}(2015)\citenamefont
  {Vinuesa}, \citenamefont {Schlatter}, \citenamefont {Malm}, \citenamefont
  {Mavriplis},\ and\ \citenamefont {Henningson}}]{vin15}%
  \BibitemOpen
  \bibfield  {author} {\bibinfo {author} {\bibfnamefont {R.}~\bibnamefont
  {Vinuesa}}, \bibinfo {author} {\bibfnamefont {P.}~\bibnamefont {Schlatter}},
  \bibinfo {author} {\bibfnamefont {J.}~\bibnamefont {Malm}}, \bibinfo {author}
  {\bibfnamefont {C.}~\bibnamefont {Mavriplis}},\ and\ \bibinfo {author}
  {\bibfnamefont {D.~S.}\ \bibnamefont {Henningson}},\ }\bibfield  {title}
  {\bibinfo {title} {Direct numerical simulation of the flow around a
  wall-mounted square cylinder under various inflow conditions},\ }\href
  {https://doi.org/10.1080/14685248.2014.989232} {\bibfield  {journal}
  {\bibinfo  {journal} {Journal of Turbulence}\ }\textbf {\bibinfo {volume}
  {16}},\ \bibinfo {pages} {555} (\bibinfo {year} {2015})},\ \Eprint
  {https://arxiv.org/abs/https://doi.org/10.1080/14685248.2014.989232}
  {https://doi.org/10.1080/14685248.2014.989232} \BibitemShut {NoStop}%
\bibitem [{\citenamefont {Zhao}\ \emph {et~al.}(2021)\citenamefont {Zhao},
  \citenamefont {Mamoon},\ and\ \citenamefont {Wu}}]{zhao_et_al_pof}%
  \BibitemOpen
  \bibfield  {author} {\bibinfo {author} {\bibfnamefont {M.}~\bibnamefont
  {Zhao}}, \bibinfo {author} {\bibfnamefont {A.-A.}\ \bibnamefont {Mamoon}},\
  and\ \bibinfo {author} {\bibfnamefont {H.}~\bibnamefont {Wu}},\ }\bibfield
  {title} {\bibinfo {title} {Numerical study of the flow past two wall-mounted
  finite-length square cylinders in tandem arrangement},\ }\href@noop {}
  {\bibfield  {journal} {\bibinfo  {journal} {Phys. Fluids}\ }\textbf {\bibinfo
  {volume} {33}},\ \bibinfo {pages} {093603} (\bibinfo {year}
  {2021})}\BibitemShut {NoStop}%
\bibitem [{\citenamefont {Germano}\ \emph {et~al.}(1991)\citenamefont
  {Germano}, \citenamefont {Piomelli}, \citenamefont {Moin},\ and\
  \citenamefont {Cabot}}]{ger91}%
  \BibitemOpen
  \bibfield  {author} {\bibinfo {author} {\bibfnamefont {M.}~\bibnamefont
  {Germano}}, \bibinfo {author} {\bibfnamefont {U.}~\bibnamefont {Piomelli}},
  \bibinfo {author} {\bibfnamefont {P.}~\bibnamefont {Moin}},\ and\ \bibinfo
  {author} {\bibfnamefont {W.~H.}\ \bibnamefont {Cabot}},\ }\bibfield  {title}
  {\bibinfo {title} {A dynamic subgrid‐scale eddy viscosity model},\ }\href
  {https://doi.org/10.1063/1.857955} {\bibfield  {journal} {\bibinfo  {journal}
  {Physics of Fluids A: Fluid Dynamics}\ }\textbf {\bibinfo {volume} {3}},\
  \bibinfo {pages} {1760} (\bibinfo {year} {1991})},\ \Eprint
  {https://arxiv.org/abs/https://doi.org/10.1063/1.857955}
  {https://doi.org/10.1063/1.857955} \BibitemShut {NoStop}%
\bibitem [{\citenamefont {{Michioka}}\ \emph {et~al.}(2014)\citenamefont
  {{Michioka}}, \citenamefont {{Takimoto}},\ and\ \citenamefont
  {{Sato}}}]{mic14}%
  \BibitemOpen
  \bibfield  {author} {\bibinfo {author} {\bibfnamefont {T.}~\bibnamefont
  {{Michioka}}}, \bibinfo {author} {\bibfnamefont {H.}~\bibnamefont
  {{Takimoto}}},\ and\ \bibinfo {author} {\bibfnamefont {A.}~\bibnamefont
  {{Sato}}},\ }\bibfield  {title} {\bibinfo {title} {{Large-Eddy Simulation of
  Pollutant Removal from a Three-Dimensional Street Canyon}},\ }\href
  {https://doi.org/10.1007/s10546-013-9870-6} {\bibfield  {journal} {\bibinfo
  {journal} {Boundary-Layer Meteorology}\ }\textbf {\bibinfo {volume} {150}},\
  \bibinfo {pages} {259} (\bibinfo {year} {2014})}\BibitemShut {NoStop}%
\bibitem [{\citenamefont {Fischer}\ \emph {et~al.}(2008)\citenamefont
  {Fischer}, \citenamefont {Lottes},\ and\ \citenamefont {Kerkemeier}}]{fis08}%
  \BibitemOpen
  \bibfield  {author} {\bibinfo {author} {\bibfnamefont {P.}~\bibnamefont
  {Fischer}}, \bibinfo {author} {\bibfnamefont {J.}~\bibnamefont {Lottes}},\
  and\ \bibinfo {author} {\bibfnamefont {S.}~\bibnamefont {Kerkemeier}},\
  }\href@noop {} {\bibinfo {title} {{NEK5000: open source spectral element CFD
  solver}}} (\bibinfo {year} {2008}),\ \bibinfo {note} {available from:
  \url{http://nek5000.mcs.anl.gov}}\BibitemShut {NoStop}%
\bibitem [{\citenamefont {Patera}(1984)}]{pat84}%
  \BibitemOpen
  \bibfield  {author} {\bibinfo {author} {\bibfnamefont {A.~T.}\ \bibnamefont
  {Patera}},\ }\bibfield  {title} {\bibinfo {title} {A spectral element method
  for fluid dynamics: Laminar flow in a channel expansion},\ }\href
  {https://doi.org/https://doi.org/10.1016/0021-9991(84)90128-1} {\bibfield
  {journal} {\bibinfo  {journal} {Journal of Computational Physics}\ }\textbf
  {\bibinfo {volume} {54}},\ \bibinfo {pages} {468} (\bibinfo {year}
  {1984})}\BibitemShut {NoStop}%
\bibitem [{\citenamefont {Varghese}\ \emph {et~al.}(2007)\citenamefont
  {Varghese}, \citenamefont {Frankel},\ and\ \citenamefont
  {Fischer}}]{varghese_frankel_fischer_2007}%
  \BibitemOpen
  \bibfield  {author} {\bibinfo {author} {\bibfnamefont {S.~S.}\ \bibnamefont
  {Varghese}}, \bibinfo {author} {\bibfnamefont {S.~H.}\ \bibnamefont
  {Frankel}},\ and\ \bibinfo {author} {\bibfnamefont {P.~F.}\ \bibnamefont
  {Fischer}},\ }\bibfield  {title} {\bibinfo {title} {{Direct numerical
  simulation of stenotic flows. Part 1. Steady flow}},\ }\href@noop {}
  {\bibfield  {journal} {\bibinfo  {journal} {J. Fluid Mech.}\ }\textbf
  {\bibinfo {volume} {582}},\ \bibinfo {pages} {253} (\bibinfo {year}
  {2007})}\BibitemShut {NoStop}%
\bibitem [{\citenamefont {Noorani}\ \emph {et~al.}(2016)\citenamefont
  {Noorani}, \citenamefont {Vinuesa}, \citenamefont {Brandt},\ and\
  \citenamefont {Schlatter}}]{duct_ref}%
  \BibitemOpen
  \bibfield  {author} {\bibinfo {author} {\bibfnamefont {A.}~\bibnamefont
  {Noorani}}, \bibinfo {author} {\bibfnamefont {R.}~\bibnamefont {Vinuesa}},
  \bibinfo {author} {\bibfnamefont {L.}~\bibnamefont {Brandt}},\ and\ \bibinfo
  {author} {\bibfnamefont {P.}~\bibnamefont {Schlatter}},\ }\bibfield  {title}
  {\bibinfo {title} {Aspect ratio effect on particle transport in turbulent
  duct flows},\ }\href@noop {} {\bibfield  {journal} {\bibinfo  {journal}
  {Phys. Fluids}\ }\textbf {\bibinfo {volume} {28}},\ \bibinfo {pages} {115103}
  (\bibinfo {year} {2016})}\BibitemShut {NoStop}%
\bibitem [{\citenamefont {Vinuesa}\ \emph {et~al.}(2018)\citenamefont
  {Vinuesa}, \citenamefont {Schlatter},\ and\ \citenamefont {Nagib}}]{vin18}%
  \BibitemOpen
  \bibfield  {author} {\bibinfo {author} {\bibfnamefont {R.}~\bibnamefont
  {Vinuesa}}, \bibinfo {author} {\bibfnamefont {P.}~\bibnamefont {Schlatter}},\
  and\ \bibinfo {author} {\bibfnamefont {H.}~\bibnamefont {Nagib}},\ }\bibfield
   {title} {\bibinfo {title} {Secondary flow in turbulent ducts with increasing
  aspect ratio},\ }\bibfield  {journal} {\bibinfo  {journal} {Physical Review
  Fluids}\ }\textbf {\bibinfo {volume} {3}},\ \href
  {https://doi.org/10.1103/PhysRevFluids.3.054606}
  {10.1103/PhysRevFluids.3.054606} (\bibinfo {year} {2018})\BibitemShut
  {NoStop}%
\bibitem [{\citenamefont {Abreu}\ \emph {et~al.}(2020)\citenamefont {Abreu},
  \citenamefont {Cavalieri}, \citenamefont {Schlatter}, \citenamefont
  {Vinuesa},\ and\ \citenamefont {Henningson}}]{pipe_ref}%
  \BibitemOpen
  \bibfield  {author} {\bibinfo {author} {\bibfnamefont {L.~I.}\ \bibnamefont
  {Abreu}}, \bibinfo {author} {\bibfnamefont {A.~V.~G.}\ \bibnamefont
  {Cavalieri}}, \bibinfo {author} {\bibfnamefont {P.}~\bibnamefont
  {Schlatter}}, \bibinfo {author} {\bibfnamefont {R.}~\bibnamefont {Vinuesa}},\
  and\ \bibinfo {author} {\bibfnamefont {D.~S.}\ \bibnamefont {Henningson}},\
  }\bibfield  {title} {\bibinfo {title} {Spectral proper orthogonal
  decomposition and resolvent analysis of near-wall coherent structures in
  turbulent pipe flows},\ }\href@noop {} {\bibfield  {journal} {\bibinfo
  {journal} {J. Fluid Mech.}\ }\textbf {\bibinfo {volume} {900}},\ \bibinfo
  {pages} {A11} (\bibinfo {year} {2020})}\BibitemShut {NoStop}%
\bibitem [{\citenamefont {Tanarro}\ \emph {et~al.}(2020)\citenamefont
  {Tanarro}, \citenamefont {Vinuesa},\ and\ \citenamefont
  {Schlatter}}]{wing_ref}%
  \BibitemOpen
  \bibfield  {author} {\bibinfo {author} {\bibfnamefont {A.}~\bibnamefont
  {Tanarro}}, \bibinfo {author} {\bibfnamefont {R.}~\bibnamefont {Vinuesa}},\
  and\ \bibinfo {author} {\bibfnamefont {P.}~\bibnamefont {Schlatter}},\
  }\bibfield  {title} {\bibinfo {title} {Effect of adverse pressure gradients
  on turbulent wing boundary layers},\ }\href@noop {} {\bibfield  {journal}
  {\bibinfo  {journal} {J. Fluid Mech.}\ }\textbf {\bibinfo {volume} {883}},\
  \bibinfo {pages} {A8} (\bibinfo {year} {2020})}\BibitemShut {NoStop}%
\bibitem [{\citenamefont {Vinuesa}\ \emph {et~al.}(2017)\citenamefont
  {Vinuesa}, \citenamefont {Fick}, \citenamefont {Negi}, \citenamefont {Marin},
  \citenamefont {Merzari},\ and\ \citenamefont {Schlatter}}]{vinuesa_toolbox}%
  \BibitemOpen
  \bibfield  {author} {\bibinfo {author} {\bibfnamefont {R.}~\bibnamefont
  {Vinuesa}}, \bibinfo {author} {\bibfnamefont {L.}~\bibnamefont {Fick}},
  \bibinfo {author} {\bibfnamefont {P.}~\bibnamefont {Negi}}, \bibinfo {author}
  {\bibfnamefont {O.}~\bibnamefont {Marin}}, \bibinfo {author} {\bibfnamefont
  {E.}~\bibnamefont {Merzari}},\ and\ \bibinfo {author} {\bibfnamefont
  {P.}~\bibnamefont {Schlatter}},\ }\bibfield  {title} {\bibinfo {title}
  {Turbulence statistics in a spectral element code: a toolbox for
  high-fidelity simulations},\ }\href@noop {} {\bibfield  {journal} {\bibinfo
  {journal} {Argonne National Lab. (ANL), Argonne, IL (United States),
  ANL/MCS-TM-367}\ } (\bibinfo {year} {2017})}\BibitemShut {NoStop}%
\bibitem [{\citenamefont {Simens}\ \emph {et~al.}(2009)\citenamefont {Simens},
  \citenamefont {Jimenez}, \citenamefont {Hoyas},\ and\ \citenamefont
  {Mizuno}}]{sim09}%
  \BibitemOpen
  \bibfield  {author} {\bibinfo {author} {\bibfnamefont {M.~P.}\ \bibnamefont
  {Simens}}, \bibinfo {author} {\bibfnamefont {J.}~\bibnamefont {Jimenez}},
  \bibinfo {author} {\bibfnamefont {S.}~\bibnamefont {Hoyas}},\ and\ \bibinfo
  {author} {\bibfnamefont {Y.}~\bibnamefont {Mizuno}},\ }\bibfield  {title}
  {\bibinfo {title} {A high-resolution code for turbulent boundary layers},\
  }\href {https://doi.org/{10.1016/j.jcp.2009.02.031}} {\bibfield  {journal}
  {\bibinfo  {journal} {J. Comput. Phys.}\ }\textbf {\bibinfo {volume} {228}},\
  \bibinfo {pages} {4218} (\bibinfo {year} {2009})}\BibitemShut {NoStop}%
\bibitem [{\citenamefont {Hoyas}\ \emph {et~al.}(2022)\citenamefont {Hoyas},
  \citenamefont {Oberlack}, \citenamefont {Alc\'antara-\'Avila}, \citenamefont
  {Kraheberger},\ and\ \citenamefont {Laux}}]{hoy22}%
  \BibitemOpen
  \bibfield  {author} {\bibinfo {author} {\bibfnamefont {S.}~\bibnamefont
  {Hoyas}}, \bibinfo {author} {\bibfnamefont {M.}~\bibnamefont {Oberlack}},
  \bibinfo {author} {\bibfnamefont {F.}~\bibnamefont {Alc\'antara-\'Avila}},
  \bibinfo {author} {\bibfnamefont {S.~V.}\ \bibnamefont {Kraheberger}},\ and\
  \bibinfo {author} {\bibfnamefont {J.}~\bibnamefont {Laux}},\ }\bibfield
  {title} {\bibinfo {title} {Wall turbulence at high friction reynolds
  numbers},\ }\href {https://doi.org/10.1103/PhysRevFluids.7.014602} {\bibfield
   {journal} {\bibinfo  {journal} {Phys. Rev. Fluids}\ }\textbf {\bibinfo
  {volume} {7}},\ \bibinfo {pages} {014602} (\bibinfo {year}
  {2022})}\BibitemShut {NoStop}%
\bibitem [{\citenamefont {Canuto}\ \emph {et~al.}(2012)\citenamefont {Canuto},
  \citenamefont {Hussaini}, \citenamefont {Quarteroni}, \citenamefont
  {Thomas~Jr} \emph {et~al.}}]{can12}%
  \BibitemOpen
  \bibfield  {author} {\bibinfo {author} {\bibfnamefont {C.}~\bibnamefont
  {Canuto}}, \bibinfo {author} {\bibfnamefont {M.~Y.}\ \bibnamefont
  {Hussaini}}, \bibinfo {author} {\bibfnamefont {A.~M.}\ \bibnamefont
  {Quarteroni}}, \bibinfo {author} {\bibfnamefont {A.}~\bibnamefont
  {Thomas~Jr}}, \emph {et~al.},\ }\href@noop {} {\emph {\bibinfo {title}
  {Spectral methods in fluid dynamics}}}\ (\bibinfo  {publisher} {Springer
  Science \& Business Media},\ \bibinfo {year} {2012})\BibitemShut {NoStop}%
\bibitem [{\citenamefont {Lluesma-Rodríguez}\ \emph
  {et~al.}(2021)\citenamefont {Lluesma-Rodríguez}, \citenamefont {Álcantara
  Ávila}, \citenamefont {Pérez-Quiles},\ and\ \citenamefont
  {Hoyas}}]{llu21c}%
  \BibitemOpen
  \bibfield  {author} {\bibinfo {author} {\bibfnamefont {F.}~\bibnamefont
  {Lluesma-Rodríguez}}, \bibinfo {author} {\bibfnamefont {F.}~\bibnamefont
  {Álcantara Ávila}}, \bibinfo {author} {\bibfnamefont {M.}~\bibnamefont
  {Pérez-Quiles}},\ and\ \bibinfo {author} {\bibfnamefont {S.}~\bibnamefont
  {Hoyas}},\ }\bibfield  {title} {\bibinfo {title} {A code for simulating heat
  transfer in turbulent channel flow},\ }\bibfield  {journal} {\bibinfo
  {journal} {Mathematics}\ }\textbf {\bibinfo {volume} {9}},\ \href
  {https://doi.org/10.3390/math9070756} {10.3390/math9070756} (\bibinfo {year}
  {2021})\BibitemShut {NoStop}%
\bibitem [{\citenamefont {Negi}\ \emph {et~al.}(2018)\citenamefont {Negi},
  \citenamefont {Vinuesa}, \citenamefont {Hanifi}, \citenamefont {Schlatter},\
  and\ \citenamefont {Henningson}}]{neg18}%
  \BibitemOpen
  \bibfield  {author} {\bibinfo {author} {\bibfnamefont {P.}~\bibnamefont
  {Negi}}, \bibinfo {author} {\bibfnamefont {R.}~\bibnamefont {Vinuesa}},
  \bibinfo {author} {\bibfnamefont {A.}~\bibnamefont {Hanifi}}, \bibinfo
  {author} {\bibfnamefont {P.}~\bibnamefont {Schlatter}},\ and\ \bibinfo
  {author} {\bibfnamefont {D.}~\bibnamefont {Henningson}},\ }\bibfield  {title}
  {\bibinfo {title} {Unsteady aerodynamic effects in small-amplitude pitch
  oscillations of an airfoil},\ }\href
  {https://doi.org/https://doi.org/10.1016/j.ijheatfluidflow.2018.04.009}
  {\bibfield  {journal} {\bibinfo  {journal} {International Journal of Heat and
  Fluid Flow}\ }\textbf {\bibinfo {volume} {71}},\ \bibinfo {pages} {378}
  (\bibinfo {year} {2018})}\BibitemShut {NoStop}%
\bibitem [{\citenamefont {Sini}\ \emph {et~al.}(1996)\citenamefont {Sini},
  \citenamefont {Anquetin},\ and\ \citenamefont {Mestayer}}]{sin96}%
  \BibitemOpen
  \bibfield  {author} {\bibinfo {author} {\bibfnamefont {J.-F.}\ \bibnamefont
  {Sini}}, \bibinfo {author} {\bibfnamefont {S.}~\bibnamefont {Anquetin}},\
  and\ \bibinfo {author} {\bibfnamefont {P.~G.}\ \bibnamefont {Mestayer}},\
  }\bibfield  {title} {\bibinfo {title} {Pollutant dispersion and thermal
  effects in urban street canyons},\ }\href@noop {} {\bibfield  {journal}
  {\bibinfo  {journal} {Atmospheric environment}\ }\textbf {\bibinfo {volume}
  {30}},\ \bibinfo {pages} {2659} (\bibinfo {year} {1996})}\BibitemShut
  {NoStop}%
\bibitem [{\citenamefont {Dong}\ \emph {et~al.}(2014)\citenamefont {Dong},
  \citenamefont {Karniadakis},\ and\ \citenamefont {Chryssostomidis}}]{don14}%
  \BibitemOpen
  \bibfield  {author} {\bibinfo {author} {\bibfnamefont {S.}~\bibnamefont
  {Dong}}, \bibinfo {author} {\bibfnamefont {G.}~\bibnamefont {Karniadakis}},\
  and\ \bibinfo {author} {\bibfnamefont {C.}~\bibnamefont {Chryssostomidis}},\
  }\bibfield  {title} {\bibinfo {title} {{A robust and accurate outflow
  boundary condition for incompressible flow simulations on severely-truncated
  unbounded domains}},\ }\href@noop {} {\bibfield  {journal} {\bibinfo
  {journal} {Journal of Computational Physics}\ }\textbf {\bibinfo {volume}
  {261}},\ \bibinfo {pages} {83} (\bibinfo {year} {2014})}\BibitemShut
  {NoStop}%
\bibitem [{\citenamefont {Schlatter}\ and\ \citenamefont
  {Örlü}(2012)}]{sch12}%
  \BibitemOpen
  \bibfield  {author} {\bibinfo {author} {\bibfnamefont {P.}~\bibnamefont
  {Schlatter}}\ and\ \bibinfo {author} {\bibfnamefont {R.}~\bibnamefont
  {Örlü}},\ }\bibfield  {title} {\bibinfo {title} {Turbulent boundary layers
  at moderate reynolds numbers: inflow length and tripping effects},\ }\href
  {https://doi.org/10.1017/jfm.2012.324} {\bibfield  {journal} {\bibinfo
  {journal} {Journal of Fluid Mechanics}\ }\textbf {\bibinfo {volume} {710}},\
  \bibinfo {pages} {5–34} (\bibinfo {year} {2012})}\BibitemShut {NoStop}%
\bibitem [{\citenamefont {Hosseini}\ \emph {et~al.}(2016)\citenamefont
  {Hosseini}, \citenamefont {Vinuesa}, \citenamefont {Schlatter}, \citenamefont
  {Hanifi},\ and\ \citenamefont {Henningson}}]{hosseini_et_al}%
  \BibitemOpen
  \bibfield  {author} {\bibinfo {author} {\bibfnamefont {S.~M.}\ \bibnamefont
  {Hosseini}}, \bibinfo {author} {\bibfnamefont {R.}~\bibnamefont {Vinuesa}},
  \bibinfo {author} {\bibfnamefont {P.}~\bibnamefont {Schlatter}}, \bibinfo
  {author} {\bibfnamefont {A.}~\bibnamefont {Hanifi}},\ and\ \bibinfo {author}
  {\bibfnamefont {D.~S.}\ \bibnamefont {Henningson}},\ }\bibfield  {title}
  {\bibinfo {title} {Direct numerical simulation of the flow around a wing
  section at moderate reynolds number},\ }\href@noop {} {\bibfield  {journal}
  {\bibinfo  {journal} {Int. J. Heat Fluid Flow}\ }\textbf {\bibinfo {volume}
  {61}},\ \bibinfo {pages} {117} (\bibinfo {year} {2016})}\BibitemShut
  {NoStop}%
\bibitem [{\citenamefont {Vinuesa}\ \emph {et~al.}(2016)\citenamefont
  {Vinuesa}, \citenamefont {Bobke}, \citenamefont {\"Orl\"u},\ and\
  \citenamefont {Schlatter}}]{vinuesa_pof}%
  \BibitemOpen
  \bibfield  {author} {\bibinfo {author} {\bibfnamefont {R.}~\bibnamefont
  {Vinuesa}}, \bibinfo {author} {\bibfnamefont {A.}~\bibnamefont {Bobke}},
  \bibinfo {author} {\bibfnamefont {R.}~\bibnamefont {\"Orl\"u}},\ and\
  \bibinfo {author} {\bibfnamefont {P.}~\bibnamefont {Schlatter}},\ }\bibfield
  {title} {\bibinfo {title} {On determining characteristic length scales in
  pressure-gradient turbulent boundary layers},\ }\href@noop {} {\bibfield
  {journal} {\bibinfo  {journal} {Phys. Fluids}\ }\textbf {\bibinfo {volume}
  {28}},\ \bibinfo {pages} {055101} (\bibinfo {year} {2016})}\BibitemShut
  {NoStop}%
\bibitem [{\citenamefont {Eitel-Amor}\ \emph {et~al.}(2014)\citenamefont
  {Eitel-Amor}, \citenamefont {Örlü},\ and\ \citenamefont
  {Schlatter}}]{eit14}%
  \BibitemOpen
  \bibfield  {author} {\bibinfo {author} {\bibfnamefont {G.}~\bibnamefont
  {Eitel-Amor}}, \bibinfo {author} {\bibfnamefont {R.}~\bibnamefont {Örlü}},\
  and\ \bibinfo {author} {\bibfnamefont {P.}~\bibnamefont {Schlatter}},\
  }\bibfield  {title} {\bibinfo {title} {Simulation and validation of a
  spatially evolving turbulent boundary layer up to $re_\theta=8300$},\ }\href
  {https://doi.org/https://doi.org/10.1016/j.ijheatfluidflow.2014.02.006}
  {\bibfield  {journal} {\bibinfo  {journal} {International Journal of Heat and
  Fluid Flow}\ }\textbf {\bibinfo {volume} {47}},\ \bibinfo {pages} {57}
  (\bibinfo {year} {2014})}\BibitemShut {NoStop}%
\bibitem [{\citenamefont {Pope}(2000)}]{pop00}%
  \BibitemOpen
  \bibfield  {author} {\bibinfo {author} {\bibfnamefont {S.~B.}\ \bibnamefont
  {Pope}},\ }\href@noop {} {\emph {\bibinfo {title} {Turbulent flows}}}\
  (\bibinfo  {publisher} {Cambridge University Press},\ \bibinfo {year}
  {2000})\BibitemShut {NoStop}%
\end{thebibliography}%

\end{document}